\documentclass[twocolumn, twocolappendix, float]{aastex63} 
\usepackage{amsmath}
\usepackage{mathtools}
\usepackage{enumitem}
\usepackage{soul}
\usepackage[normalem]{ulem}
\usepackage{CJK}
\usepackage{hyperref}
\hypersetup{
    colorlinks=true,
    linkcolor=blue,
    filecolor=magenta,      
    urlcolor=blue}

\newcommand{\mbh}{M_{\rm BH}}
\newcommand{\mstar}{M_{\rm *}}

\graphicspath{{./}{figures/}}
\shorttitle{TDE in dwarfs}
\shortauthors{Tan et al}

\begin{document}
\begin{CJK*}{UTF8}{gbsn}

\title{Rare Occasions: Tidal Disruption Events Rarely Power the AGNs Observed in Dwarf Galaxies}
\correspondingauthor{Joanne Tan, Guang Yang}
\email{jtan@mpa-garching.mpg.de, gyang@niaot.ac.cn}

\author[0009-0006-5896-2554]{Joanne Tan}
\affiliation{Max-Planck-Institut f\"ur Astrophysik, Karl-Schwarzschild-Stra{\ss}e 1, D-85748, Garching, Germany}
\affiliation{Department of Physics and Astronomy, Texas A\&M
  University, College Station, TX, 77843-4242 USA}
\affiliation{George P.\ and Cynthia Woods Mitchell Institute for
 Fundamental Physics and Astronomy, Texas A\&M University, College
 Station, TX, 77843-4242 USA}
 
\author[0000-0001-8835-7722]{Guang Yang (杨光)}
\affiliation{Nanjing Institute of Astronomical Optics \& Technology, Chinese Academy of Sciences, Nanjing 210042, China}
\affiliation{CAS Key Laboratory of Astronomical Optics \& Technology, Nanjing Institute of Astronomical Optics \& Technology, Nanjing 210042, China}
\affiliation{Kapteyn Astronomical Institute, University of Groningen, P.O. Box 800, 9700 AV Groningen, The Netherlands}
\affiliation{SRON Netherlands Institute for Space Research, Postbus 800, 9700 AV Groningen, The Netherlands}
\affiliation{Department of Physics and Astronomy, Texas A\&M
  University, College Station, TX, 77843-4242 USA}
\affiliation{George P.\ and Cynthia Woods Mitchell Institute for
 Fundamental Physics and Astronomy, Texas A\&M University, College
 Station, TX, 77843-4242 USA}

\author[0000-0002-1881-5908]{Jonelle L. Walsh}
\affiliation{Department of Physics and Astronomy, Texas A\&M
  University, College Station, TX, 77843-4242 USA}
\affiliation{George P.\ and Cynthia Woods Mitchell Institute for
 Fundamental Physics and Astronomy, Texas A\&M University, College
 Station, TX, 77843-4242 USA} 

\author[0000-0002-0167-2453]{W. N. Brandt}
\affiliation{Department of Astronomy and Astrophysics, 525 Davey Lab, The Pennsylvania State University, University Park, PA 16802, USA}
\affiliation{Institute for Gravitation and the Cosmos, The Pennsylvania State University, University Park, PA 16802, USA}
\affiliation{Department of Physics, 104 Davey Laboratory, The Pennsylvania State University, University Park, PA 16802, USA}

\author[0000-0002-9036-0063]{Bin Luo}
\affiliation{School of Astronomy and Space Science, Nanjing University, Nanjing, Jiangsu 210046, China}

\author[0000-0002-8686-8737]{Franz E. Bauer}
\affiliation{Instituto de Astrof\'{\i}sica, Facultad de F\'{i}sica, Pontificia Universidad Cat\'{o}lica de Chile, Casilla 306, Santiago 22, Chile}
\affiliation{Millennium Institute of Astrophysics, Nuncio Monse\~{n}or S\'{o}tero Sanz 100, Providencia, Santiago, Chile}
\affiliation{Space Science Institute, 4750 Walnut Street, Suite 205, Boulder, Colorado 80301}

\author[0000-0002-4945-5079]{Chien-Ting Chen}
\affiliation{Astrophysics Office, NASA Marshall Space Flight Center, ZP12, Huntsville, AL 35812, USA}

\author[0000-0002-0771-2153]{Mouyuan Sun}
\affiliation{Department of Astronomy, Xiamen University, Xiamen, Fujian 361005, People's Republic of China}

\author[0000-0002-1935-8104]{Yongquan Xue}
\affiliation{CAS Key Laboratory for Research in Galaxies and Cosmology, Department of Astronomy, University of Science and Technology of China, Hefei 230026, China}
\affiliation{School of Astronomy and Space Sciences, University of Science and Technology of China, Hefei 230026, China}

\begin{abstract}
Tidal disruption events (TDEs) could be an important growth channel for massive black holes in dwarf galaxies. 
Theoretical work suggests that the observed active galactic nuclei (AGNs) in dwarf galaxies are predominantly TDE-powered. 
To assess this claim, we perform variability analyses on the dwarf-hosted AGNs detected in the $7$ Ms \textit{Chandra} Deep Field-South (CDF-S) survey, with observations spanning $\approx 16$ years.
Based on the spectral energy distribution (SED) modeling with {\sc x-cigale}, we select AGNs hosted by dwarf galaxies (stellar mass below $10^{10}\ M_\odot$). 
We focus on X-ray sources with full-band detections, leading to a sample of $78$ AGNs (0.122 $\leq$ $z$ $\leq$ 3.515). 
We fit the X-ray light curves with a canonical TDE model of $t^{-5/3}$ and a constant model. 
If the former outperforms the latter in fitting quality for a source, we consider the source as a potential TDE. 
We identify five potential TDEs, constituting a small fraction of our sample. 
Using true- and false-positive rates obtained from fitting models to simulated light curves, we perform Bayesian analysis to obtain the posterior of the TDE fraction for our sample.
The posterior peaks close to zero (2.56\%), and we obtain a $2$-$\sigma$ upper limit of $9.80\%$. 
Therefore, our result indicates that the observed AGNs in dwarf galaxies are \textit{not} predominantly powered by TDEs. 
\end{abstract}

\section{Introduction}
\label{sec1: intro}
Massive black holes (BHs) are common in galactic centers \citep[e.g.,][]{kormendy13}. Stars approaching the BH can be pulled apart by the BH's tidal force, when the tidal force has surpassed the star's self-gravity \citep[e.g.,][]{Rees1988Natur.333..523R}. 
While some of the stellar materials get expelled at high velocities, the remainder can get accreted, feeding the BH at super-Eddington rates and causing a bright flare that lasts for several years. Such events are known as tidal disruption events (TDEs). 

The contribution of TDEs to overall BH growth might depend on the BH mass ($\mbh$). The dynamical models of \cite{Magorrian1999MNRAS.309..447M} estimated that TDEs can account for $\sim 10^{6}\ M_\odot$ BH growth across cosmic history, independent of galaxy luminosity. This amount of BH growth is negligible for massive BHs, whose dominant growth channel should be cold-gas accretion. However, for low-mass BHs ($\mbh \lesssim 10^{7}\ M_\odot$), the TDE-related BH growth can account for a large fraction of the total $\mbh$ (see also, e.g., \citealt{Wang2004ApJ...600..149W, Stone2016MNRAS.455..859S, greene20} for similar conclusions). Therefore, TDEs could be a crucial ingredient for the growth of the low-mass BHs hosted by dwarf galaxies (stellar mass, $\mstar \lesssim 10^{10} M_\odot$).

\cite{Zubovas2019MNRAS.483.1957Z} modeled the active galactic nucleus (AGN) duty cycle powered by TDEs in dwarf galaxies, and found the predicted TDE-powered AGN fraction ($\gtrsim 0.5\%$) is consistent with the observed AGN fraction, with the assumption that the occupation fraction in dwarf galaxies is close to unity. Their result suggests that the observed AGNs in dwarf galaxies could be predominantly driven by TDEs. This theoretical conclusion is testable from observations: one can investigate the temporal behavior of AGNs detected in dwarf galaxies and determine if they are powered by TDEs. 
We perform such a study in this work. 

The TDE flares mainly emit at UV and X-ray wavelengths \citep[e.g.,][]{Rees1988Natur.333..523R}. Compared to UV photons, X-rays are less affected by obscuration due to their high penetrating power. Therefore, X-ray observations are ideal for the search for TDEs. Indeed, X-ray facilities have successfully identified many TDE candidates. For example, 2XMM J184725.1-631724 is an ultrasoft X-ray flare discovered by \textit{XMM-Newton} \citep{Lin2011ApJ...738...52L}. The intensive multiwavelength follow-up observations strongly support the flare being a TDE \citep{Lin2018MNRAS.474.3000L}. RBS~1032 is a decaying X-ray source detected by the multi-epoch data of \textit{ROSAT} \citep{fischer98}. It also has a supersoft spectrum in all \textit{ROSAT} epochs \citep{ghosh06}.
The \textit{XMM-Newton} follow-up observations confirm that its light curve follows the $t^{-5/3}$ evolution track, a fundamental TDE signature \citep{Maksym2014ApJ...792L..29M}. From optical observations, the host galaxy of RBS~1032 is a dwarf galaxy at $z=0.024$ \citep{ghosh06}. More recently, the eROSITA telescope \citep{eROSITA_Predehl2021A&A...647A...1P} onboard the SRG observatory \citep{eROSITA_Sunyaev2021A&A...656A.132S} has discovered a sample of 13 TDEs \citep{eROSITA_Sazonov2021MNRAS.508.3820S}, which are confirmed by follow-up optical observations.

Given the advantages and successes of X-ray TDE searches, we choose to utilize the $7$~Ms \textit{Chandra} Deep Field-South (CDF-S) data (\citealt{Luo2017ApJS..228....2L}, L17 hereafter) to examine if the AGNs in dwarf galaxies are powered by TDEs, as hypothesized by \cite{Zubovas2019MNRAS.483.1957Z}. The $7$~Ms CDF-S is currently the deepest X-ray survey (see \citealt{Xue2017NewAR..79...59X} for a review), enabling the detection of faint AGNs in dwarf galaxies.
The CDF-S observations span a temporal baseline of $\approx 16$ years. This long baseline significantly benefits our TDE search, because TDE light curves 
have stronger flux changes toward longer timescales. Short-term ($\lesssim$~ a few months) variability analyses may not be able to effectively differentiate TDEs from the variability of ordinary AGNs \citep[e.g.,][]{Maksym2014ApJ...792L..29M, Yang2016ApJ...831..145Y}. Another advantage of the long baseline is the increased probability of a TDE occurrence. 
The CDF-S is also accompanied by deep observations from UV to radio wavelengths \citep[e.g.,][]{elbaz11, Guo2013ApJS..207...24G}. The multiwavelength data are critical for reliably estimating $\mstar$ and thereby selecting for dwarf galaxies for our study. 

This paper is structured as follows. In Section \ref{sec2: data select}, we describe the observations, source selection, and spectral energy distribution (SED) fitting of the sources. 
In Section \ref{sec3: analysis}, we perform variability analyses and identify TDE candidates. 
Next, we constrain the TDE fraction in Section \ref{sec4: sim}. 
In Section \ref{sec5: disc}, 
we compare the ratio of fluxes between pairs of epochs for each source and examine the distribution of the effective power-law photon index, $\Gamma$.
These diagnostics serve to identify the model-independent changes in fluxes and to determine whether the sources have $\Gamma$ values typical of TDEs, respectively. 
Lastly, we summarize our results and discuss future prospects in Section \ref{sec6: sumz}. 

Throughout this paper, we assume a cosmology with $H_0$ = $70$ km s$^{-1}$  Mpc$^{-1}$, $\Omega_{\rm M}$ = $0.3$, and $\Omega_{\Lambda}$ = $0.7$. 
We use AB magnitudes and quote uncertainties at the $1$-$\sigma$ (68\%) confidence level, unless stated otherwise. We define ``AGNs" to include both ordinary AGNs and TDE-powered AGNs; we emphasize that ``ordinary AGNs" refer only to AGNs powered by typical gas accretion.

\section{Data and Source Selection}
\label{sec2: data select}
\subsection{Observations}
\label{subsec2-1: obs}

This work, based on the $7$~Ms CDF-S X-ray data, employs a list of Chandra datasets, obtained by the Chandra X-ray Observatory, contained in the Chandra Data Collection (CDC) 270~\dataset[doi:10.25574/cdc.270]{https://doi.org/10.25574/cdc.270}. 
There are a total of $102$ observations, which were taken from October 1999 to March 2016. 
The median exposure time of the $102$ observations is $\approx 56$ ks. 
All of the $102$ CDF-S observations utilized the \textit{Chandra} Advanced CCD Imaging Spectrometer imaging array \citep[ACIS-I,][]{Garmire2003SPIE.4851...28G}. 
More details about the observations can be found in Section 2.1 of L17. 
We use the {\sc ACIS Extract} \citep{Broos2010ApJ...714.1582B} data products from L17 (see Section 2.2 of L17 for details on data reduction). 

Most CDF-S sources have low signal-to-noise (S/N) in single observations. Thus, to improve the S/N of the sources, we bin data from observations made around similar times to form several ``epochs''. 
From the $102$ observations, we compile five epochs for each source, and each of these epochs consists of observations that sum up to $\sim$ $1$--$2$~Ms of exposure time. 
Similar to \citet[][]{Yang2016ApJ...831..145Y}, the bins are chosen such that as much data as possible are included in the shortest span of time possible. 
This is to minimize the variability effects within each epoch. 
The epochs' details are listed in Table \ref{tab: epoch}.

\begin{deluxetable}{cccccc}[h!]
\tablenum{1}
\tablecaption{Epochs formed from 102 observations\label{tab: epoch}}
\tablewidth{0pt}
\tablehead{
\colhead{Epoch} & \colhead{Start date} & \colhead{End date} & \colhead{$W_{\rm bin}$ (yr)$^a$} & \colhead{$T_{\rm exp}$ (Ms)$^b$} & \colhead{$N_{\rm obs}$ $^c$}
}
\startdata 
1 & 1999 Oct & 2000 Dec & 1.19 & 0.931 & 11 \\
2 & 2007 Sep & 2007 Nov & 0.12 & 0.960 & 12 \\
3 & 2010 Mar & 2010 Jul & 0.35 & 1.956 & 31 \\
4 & 2014 Jun & 2015 Jan & 0.57 & 1.854 & 31 \\
5 & 2015 Mar & 2016 Mar & 1.02 & 1.027 & 17 \\
\enddata
\tablecomments{\\ $^a$ Bin width, i.e., the time span between the start and end dates.
\\ $^b$ Total exposure time of all observations in an epoch.
\\ $^c$ Number of observations in each epoch.}
\end{deluxetable}

\subsection{Source selection}
\label{subsec2-2: source selection}
We first select our sources from the main catalog of the $7$~Ms CDF-S survey (hereafter the L17 catalog). 
From these, we narrow down to sources with a counterpart in the CANDELS multi-wavelength catalog in the GOODS-South Field \citep[hereafter the CANDELS catalog]{Guo2013ApJS..207...24G}, adopting the counterpart-matching results from the L17 catalog. 
We find a subset of $674$ sources that are in both catalogs, covering a sky region of $\approx 170$ arcmin$^2$.

Next, we remove all sources that are classified as ``Galactic star'' (identified spectroscopically) in the L17 catalog and retain only sources that are classified as ``AGN" or ``Galaxy". At this stage, we have $670$ sources. 
We then also remove two sources that are confirmed X-ray transients, i.e. CDF-S XT1 \citep{Bauer2017MNRAS.467.4841B} and CDF-S XT2 \citep{Xue2019Natur.568..198X}, as these sources are unlikely to have AGNs and we are focusing only on galaxies that host AGNs to search for potential TDEs. 

From the $668$ sources, we filter for sources with full-band ($0.5$--$7.0$ keV) detections. We use full-band photometry to construct light curves and search for TDE signatures as it encompasses both the soft and hard bands, making it the widest available band and yielding the largest number of counts for each source.
This gives us $591$ sources.
Lastly, to select low-mass sources, we perform multi-wavelength SED fitting for these $591$ sources using {\sc x-cigale} (\citealt{Boquien2019A&A...622A.103B, Yang2020MNRAS.491..740Y}; see Section \ref{subsec2-3: SED fitting} for details). 
We define low-mass sources to be those with $M_{*} \leq 10^{10}\ M_{\odot}$, as determined by {\sc x-cigale}. This criterion gives us $144$ low-mass sources. In this work, we are focusing on AGNs hosted by dwarf galaxies; thus, we select for low-mass sources that L17 have classified as ``AGN". Ultimately, we obtain $78$ X-ray detected AGNs in low-mass galaxies with redshift ($z$, from L17) between $0.122$ and $3.515$.
The properties of our sample are tabulated in Table \ref{tab: data}.

\begin{deluxetable*}{crrcccccccccr}
\tabletypesize{\scriptsize}
\tablenum{2}
\tablecaption{\label{tab: data} Properties of low-mass AGNs}
\tablewidth{0pt}
\tablehead{
\colhead{Source \#}  & \colhead{ID$_{\rm L17}$} & \colhead{ID$_{\rm G13}$} & \colhead{RA ($^{\circ}$)} & \colhead{Dec ($^{\circ}$)} & \colhead{$z$} & \colhead{$z$-$type$} & \colhead{$\theta$ ($'$)} & \colhead{$C_{\rm L17}$} & \colhead{$m_{\rm F160W}$} & \colhead{log($M_*$)}
& \colhead{log ($M_{*, err}$)} & \colhead{$\Delta \mathrm{AIC}$} \\[-2.5ex] 
\colhead{(1)} & \colhead{(2)} & \colhead{(3)} & \colhead{(4)} & \colhead{(5)} & \colhead{(6)} & \colhead{(7)} & \colhead{(8)} & \colhead{(9)} & \colhead{(10)} & \colhead{(11)} & \colhead{(12)} & \colhead{(13)}
}  
\startdata
1  &  716  &  24716  &  53.15883  &  -27.66244  &  0.84  &  s  &  8.89  &  9399.8  &  20.00  &  9.59  &  0.63  &  1086.78  \\  
2  &  567  &  18955  &  53.12590  &  -27.75126  &  0.74  &  s  &  3.31  &  7393.2  &  21.10  &  9.87  &  0.11  &  1262.95  \\  
3  &  102  &  21937  &  53.00658  &  -27.72416  &  2.73  &  s  &  7.68  &  3250.2  &  22.60  &  9.76  &  0.60  &  155.74  \\  
4  &  252  &  6599  &  53.05835  &  -27.85018  &  0.12  &  s  &  4.12  &  3062.1  &  20.72  &  8.98  &  0.12  &  1347.90  \\  
5  &  449  &  26637  &  53.10485  &  -27.70521  &  1.62  &  s  &  6.09  &  2514.1  &  21.43  &  9.99  &  0.39  &  48.32  \\  
6  &  704  &  23708  &  53.15608  &  -27.66668  &  0.66  &  s  &  8.60  &  1702.6  &  21.13  &  9.89  &  0.14  &  173.19  \\  
7  &  788  &  14587  &  53.17848  &  -27.78402  &  3.19  &  s  &  3.48  &  1036.8  &  23.75  &  9.31  &  0.28  &  76.29  \\  
8  &  340  &  12338  &  53.07914  &  -27.79873  &  0.67  &  s  &  2.10  &  998.4  &  22.77  &  9.29  &  0.02  &  238.63  \\  
9  &  243  &  10323  &  53.05583  &  -27.81552  &  1.45  &  p  &  3.34  &  321.2  &  25.56  &  9.55  &  0.20  &  35.46  \\  
10  &  326  &  3643  &  53.07600  &  -27.87814  &  2.80  &  s  &  4.86  &  317.5  &  23.52  &  9.83  &  0.12  &  -6.25  \\
$\vdots$ & $\vdots$ & $\vdots$ & $\vdots$ & $\vdots$ & $\vdots$ & $\vdots$ & $\vdots$ & $\vdots$ & $\vdots$ & $\vdots$ & $\vdots$ & $\vdots$  \\
\enddata
\tablecomments{Column (1): source sequence number, assigned in order of descending net counts from L17. Columns (2) and (3): source ID in the L17 catalog and the CANDELS catalog, respectively. 
Columns (4) and (5): right ascension and declination in J2000 coordinates, respectively, taken from L17. Column (6): adopted redshift from L17. Column (7): type of redshift, with ``s" meaning spectroscopic redshift and ``p" referring to photometric redshift. 
Column (8): off-axis angle in units of arcminutes. Column (9): source's net counts in the full band from the {\sc ACIS Extract} data products of the L17 catalog. Column (10): AB magnitude in the Hubble Space Telescope Wide Field Camera 3 (HST/WFC3) F160W filter, converted from the flux in the same filter given in the G13 catalog. Column (11) and (12): log stellar mass in units of solar masses and its corresponding uncertainty, measured by {\sc x-cigale}.  
Column (13): difference in the Akaike Information Criterion (AIC) between the two models described in Section \ref{sec3: analysis}. 
The full table of $78$ sources is available online in a machine-readable format.
}
\end{deluxetable*}

\subsection{SED fitting with {\sc x-cigale}}
\label{subsec2-3: SED fitting}
\showboxdepth=\maxdimen
\showboxbreadth=\maxdimen
In order to measure $M_{*}$, we utilize {\sc x-cigale}\footnote{\href{https://gitlab.lam.fr/gyang/cigale/tree/xray}{https://gitlab.lam.fr/gyang/cigale/tree/xray}} \citep{Boquien2019A&A...622A.103B, Yang2020MNRAS.491..740Y}, an SED-modelling tool with the capability to build AGN and galaxy SEDs in the X-ray to IR wavelength range.
We obtain the bandpass fluxes of $17$ filters from the UV to mid-IR wavelengths from the CANDELS catalog; more details about the filters and data used can be found in \citealt[][]{Guo2013ApJS..207...24G} (G13 hereafter).
We also use X-ray fluxes from L17 in the hard band ($2$--$7$ keV) or in the full band. We prioritize fluxes in the hard band, as it is the least obscured among the three available bands and {\sc x-cigale} requires using unobscured data for the SED fitting. For sources that are not detected in the hard band, we use their full-band fluxes instead.
We use the hard-band detected fluxes for $358$ sources ($60.6\%$) and full-band detected fluxes for the remaining $233$ sources ($39.4\%$).

The {\sc x-cigale} modules we use are listed in Table \ref{tab: xcigale params}, along with their respective fitting parameters.
{\sc x-cigale} requires the assumption of a parametric star formation history (SFH) in order to model the SED of 
a galaxy. We adopt the delayed-$\tau$ SFH model. 
The stellar spectrum is computed from the SFH and the \citet{Bruzual2003MNRAS.344.1000B} single stellar population (SSP) models, assuming a \citet{Chabrier2003ApJ...586L.133C} initial mass function (IMF) and a metallicity of $Z = 0.02$ (solar). For the computation of nebular emission, we adopt the {\sc nebular} module, which considers both continuum and line nebular emission.

To compute galactic dust attenuation, we adopt {\sc dustatt\_modified\_starburst}, a module based on the \citet{Calzetti2000ApJ...533..682C} starburst attenuation, and extended with the \citet{Leitherer2002ApJS..140..303L} curve between the Lyman break and $150$~nm. We fix the color excess ratio between the old and young stellar populations to unity, 
and vary the color excess of starlight for both the young and old populations, as listed in Table \ref{tab: xcigale params}. For galactic dust emission, we adopt the model from \citet{Dale2014ApJ...784...83D} with d$M_{\mathrm dust}$ $\propto$ $U^{-\alpha}$d$U$, in which $M_{\mathrm dust}$ is the dust mass, $U$ is radiation-field intensity, and $\alpha$ is the slope. We only vary $\alpha$ and use the values of $1.0$, $1.5$, $2.0$, and $2.5$.

We adopt the {\sc skirtor} module \citep{Stalevski2012MNRAS.420.2756S, Stalevski2016MNRAS.458.2288S} to compute the AGN UV-to-IR emission. The three free parameters for the {\sc skirtor} module are listed in Table \ref{tab: xcigale params}. We allow the AGN fraction to vary since the AGN contribution may not be dominant for the sources in our sample. We set the viewing angle, $\theta$, to 30$^{\circ}$ and 70$^{\circ}$, which are the 
typical $\theta$ values for spectroscopic type 1 and type 2 AGNs, respectively 
\citep[e.g.,][]{Yang2020MNRAS.491..740Y, RamosPadilla2022MNRAS.510..687R}.
Another free parameter is the color excess of the polar dust.
We fix the other {\sc skirtor} parameters at their default values. For the X-ray module, we adopt the default values for the parameters. 

To summarize, the parameters listed in Table \ref{tab: xcigale params} are the only ones allowed to vary, 
while the parameters not listed in Table \ref{tab: xcigale params} are fixed to their default values. 
For the fitting, we utilize the Bayesian-like analysis option supported by {\sc x-cigale}, which provides the likelihood-weighted values of physical properties calculated, such as $M_*$. 
We obtain a total of $144$ sources with 
$M_*$ $\leq$ $10^{10}$ $M_{\odot}$, while the remaining $447$ sources have $M_*$ $>$ $10^{10}$ $M_{\odot}$. 
Among the low-mass sources, $78$ of them are classified as AGNs and the remaining $66$ are classified as normal galaxies.
This SED-fitting procedure helps us to fulfill the final criterion of our sample selection as described in Section \ref{subsec2-2: source selection}. 

\begin{figure}[ht!]
\includegraphics[width=\linewidth]{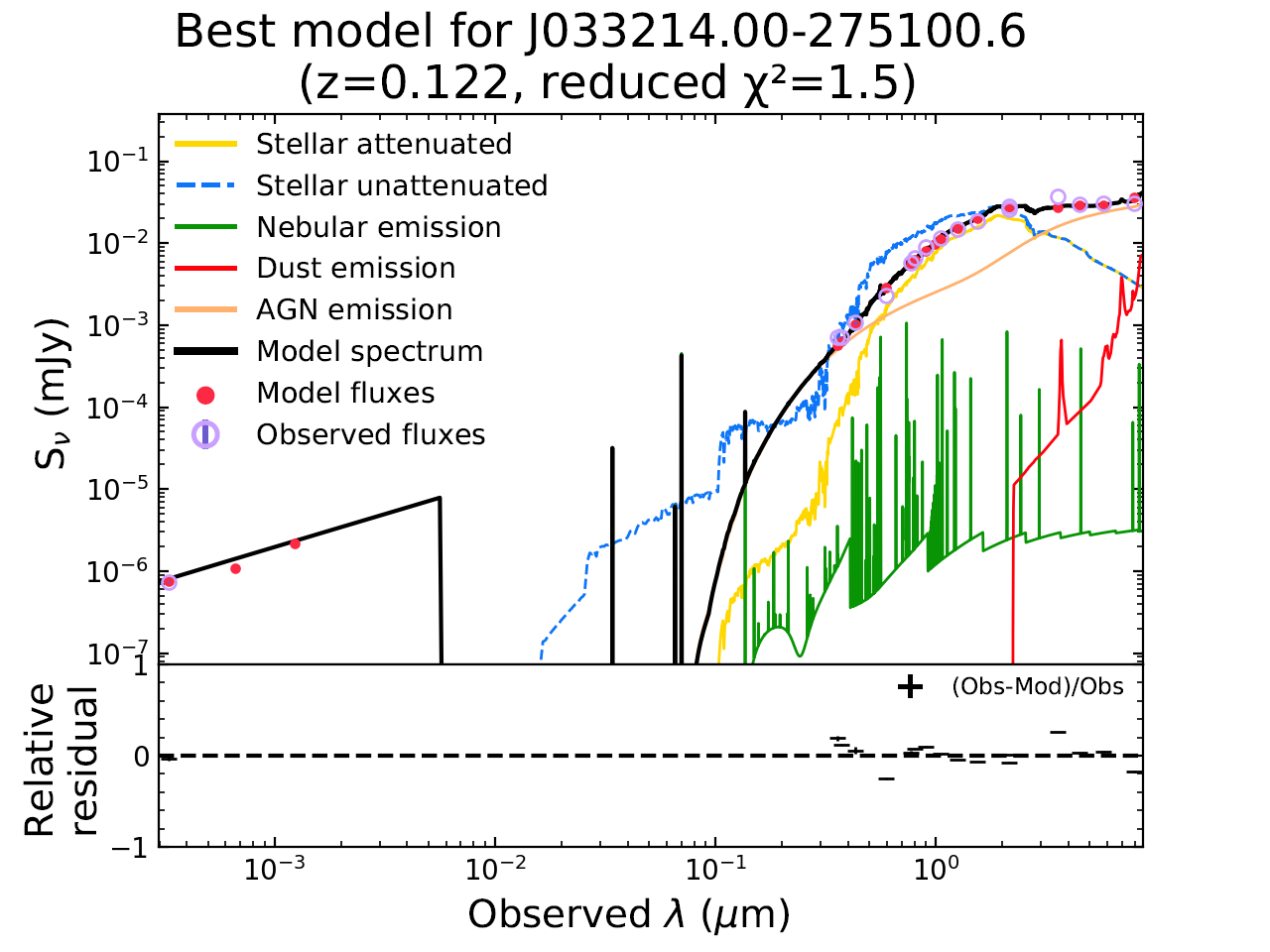} 
\includegraphics[width=\linewidth]{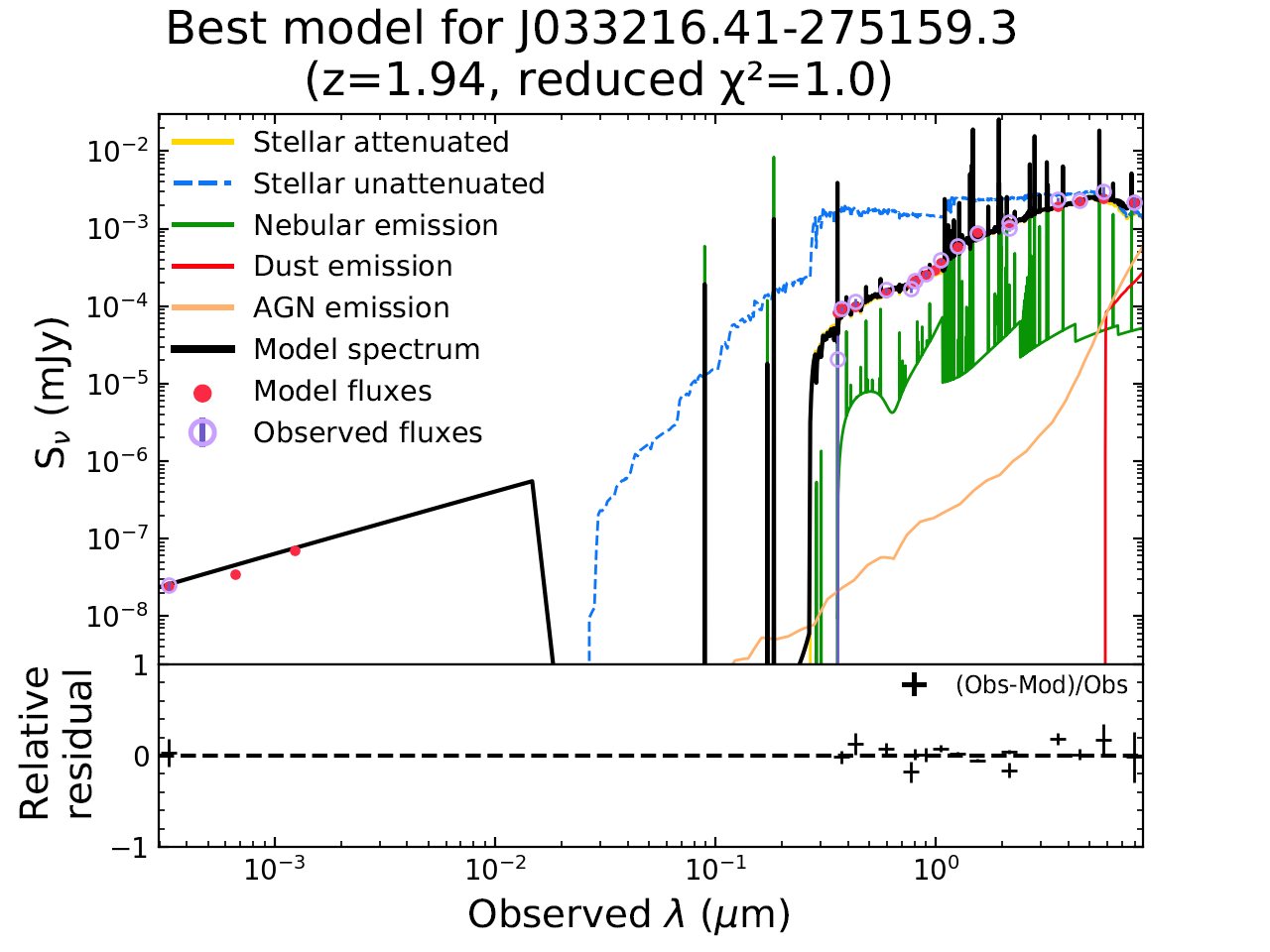} 
\caption{\label{fig:xcgiale} Examples of the SED fitting using {\sc x-cigale} for 
sources \#$4$ (top) and \#$13$ (bottom). The fitting residuals ($1$ - model flux/observed flux) are shown at the bottom of each panel.}
\end{figure}

\begin{figure}[ht!]
\includegraphics[width=\columnwidth]{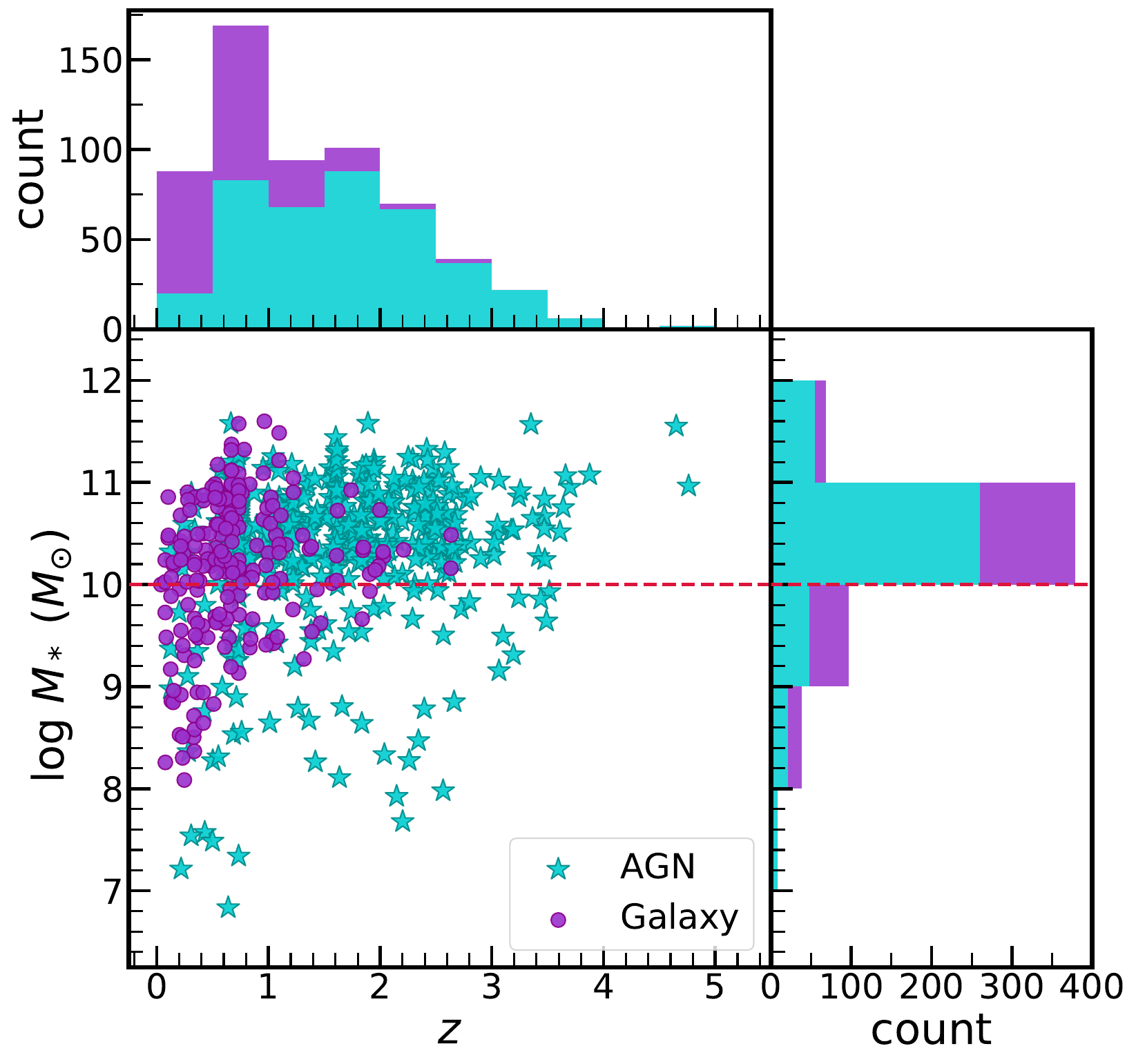}
\caption{\label{fig:mar hist} 
The stellar masses of the $591$ sources fit by {\sc x-cigale} are shown as a function of redshift.
In the scatter plot, the cyan-colored stars indicate both the low- ($M_*$ $\leq$ $10^{10}$ $M_{\odot}$) and high-mass ($M_*$ $>$ $10^{10}$ $M_{\odot}$) AGNs, while the purple-colored dots indicate sources classified as galaxies, both in the low- and high-mass ranges. 
Above the scatter plot is a stacked redshift histogram and to the right of the scatter plot is a stacked histogram of $M_*$. In both panels, the cyan histogram represents AGNs in both mass ranges and the purple histogram includes both low- and high-mass galaxies. The red dashed lines denote the separation between low-mass and high-mass sources, at $M_* =$ $10^{10}$ $M_{\odot}$.}
\end{figure}

\begin{figure}[ht!]
\includegraphics[width=\columnwidth]{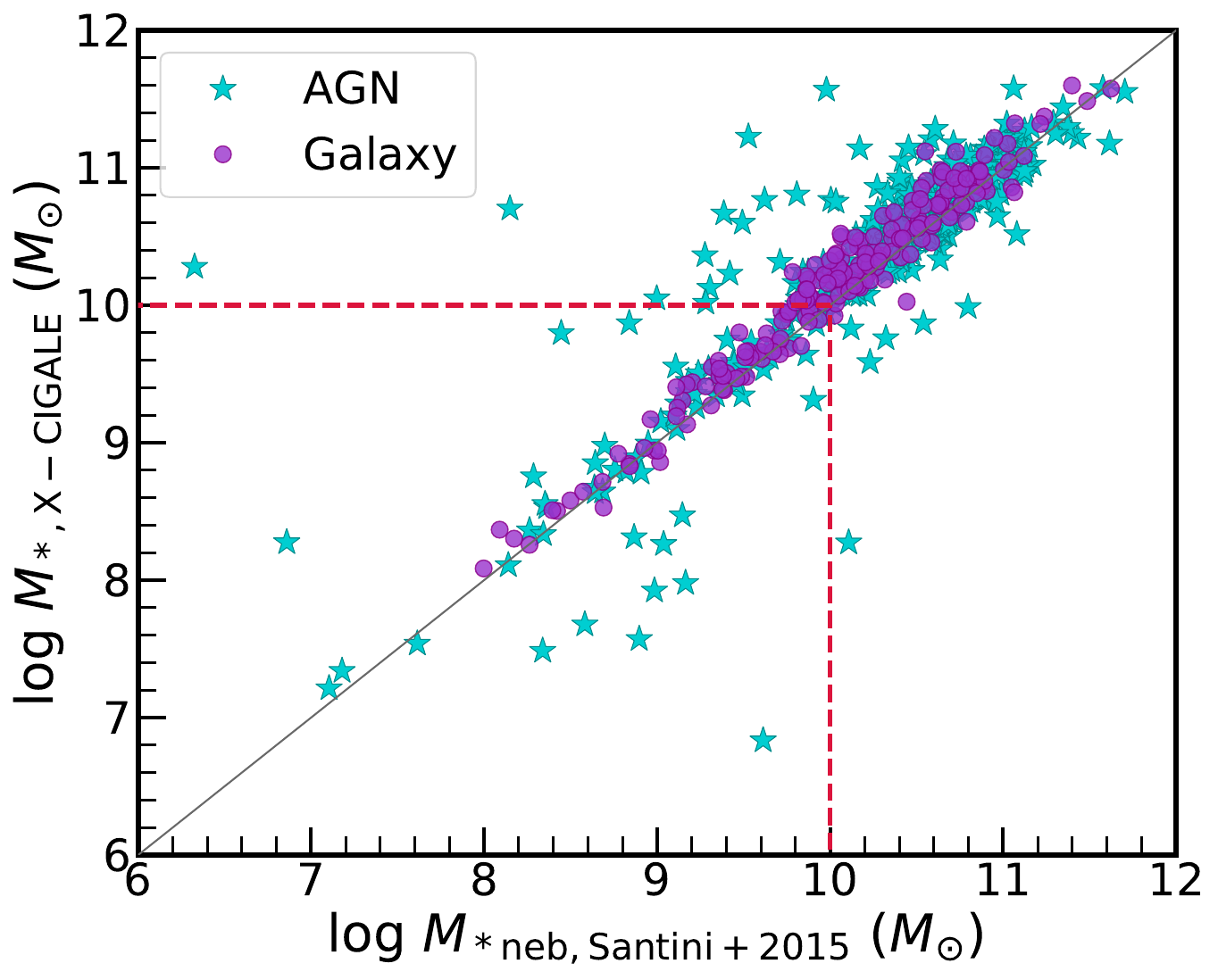}
\caption{\label{fig:mass compare} The comparison between the stellar mass from \citet{Santini2015ApJ...801...97S} and the stellar mass as fitted by {\sc x-cigale}. The figure shows a good agreement between the two sets of stellar masses for galaxies. The scatter in the stellar mass comparison for the AGNs implies that the inclusion of an AGN component (in {\sc x-cigale} but not in \citet{Santini2015ApJ...801...97S}) does affect mass estimation and is necessary for an accurate mass estimation. The red dashed lines denote the separation between low-mass and high-mass sources, at $M_* =$ $10^{10}$ $M_{\odot}$.}
\end{figure}

Figure \ref{fig:xcgiale} shows the best-fit SEDs for two sources in our sample 
and Figure \ref{fig:mar hist} shows a scatter plot of the stellar mass for the $591$ sources fitted by {\sc x-cigale} as a function of 
$z$.
This study focuses on the subset of sources marked by cyan-colored stars located below the red dashed line, which are the $78$ low-mass AGNs that fulfill all our selection criteria (see Section \ref{subsec2-2: source selection}). In Figure \ref{fig:mass compare}, we compare the stellar masses measured by {\sc x-cigale} to the stellar masses from \citet{Santini2015ApJ...801...97S}.\footnote{\citet{Santini2015ApJ...801...97S} provide stellar-mass estimates from different fitting methods carried out by various research groups. We adopt the median stellar mass from the estimates that include nebular emission in the model spectra, as it provides a relatively robust estimate, and because our {\sc x-cigale} fitting also considers nebular emission.} 
When running {\sc x-cigale} we include an AGN component, whereas the stellar masses given by \citet{Santini2015ApJ...801...97S} were estimated without considering AGN emission. As can be seen in Figure \ref{fig:mass compare}, the masses for the galaxy sample, denoted by purple dots, show good agreement between our and \citet{Santini2015ApJ...801...97S} measurements. This agreement implies that we have run {\sc x-cigale} robustly. Conversely, we can see a significant scatter in the stellar mass comparison for the AGN sample, denoted by cyan stars. This shows that the inclusion of an AGN component does affect the estimation of stellar mass and the AGN component is necessary when performing SED-fitting. Hence, for our sample, we adopt the stellar mass measured by {\sc x-cigale}.

\begin{deluxetable*}{lcc}
\tabletypesize{\scriptsize}
\tablenum{3}
\tablecaption{\label{tab: xcigale params} {\sc x-cigale} Parameters} 
\tablewidth{0pt}
\tablehead{
\colhead{Module} & \colhead{Parameters} & \colhead{Values}
}
\startdata
Star formation history: & $e$-folding time, $\tau$ (Gyr)  & 0.1, 0.5, 1, 2, 5 \\
Delayed model, SFR $\propto$ $t$ $\times$ exp(-$t/\tau$) & Stellar Age, $t$ (Gyr)  & 0.5, 1, 3, 5, 7 \\
 & & \\
Simple stellar population: \citet{Bruzual2003MNRAS.344.1000B} & Initial mass function  & \citet{Chabrier2003ApJ...586L.133C} \\
 & & \\
Galactic dust attenuation: & $E(B - V)$ ratio between the old and young populations  &  1 \\
\citet{Calzetti2000ApJ...533..682C} \& \citet{Leitherer2002ApJS..140..303L} & $E(B - V)$ of starlight  & 0, 0.01, 0.02, 0.05, \\
& & 0.1--0.5 (steps of 0.1), 0.7, 0.9 \\
 & & \\
Galactic dust emission: \citet{Dale2014ApJ...784...83D} & $\alpha$ slope in d$M_{\rm dust}$ $\propto$ $U^{-\alpha}$d$U$ & 1.0, 1.5, 2.0, 2.5 \\
 & & \\
AGN (UV-to-IR): & Viewing angle $\theta$ (face on: $\theta$=0$^{\circ}$, edge on: $\theta$=90$^{\circ}$) & 30$^{\circ}$ (type 1), 70$^{\circ}$ (type 2) \\ 
{\sc skirtor} \citep{Stalevski2012MNRAS.420.2756S, Stalevski2016MNRAS.458.2288S}  & AGN fraction in total IR luminosity, $f_{\rm AGN}$  & 0, 0.01, 0.1-–0.9 (steps of 0.1), 0.99 \\
 & $E(B - V)$ of polar dust  & 0, 0.05, 0.1, 0.15, 0.2, 0.3 \\
  & & \\
X-ray: & Maximum deviation from the $\alpha_{\rm ox}$-$L_{\rm 2500 \text{\r{A}}}$ relation & 0.2 \\
  & & \\
\enddata
\tablecomments{X-CIGALE parameters not included here are set to their default values.}
\end{deluxetable*}

\section{Data Processing and Analysis}
\label{sec3: analysis}
\subsection{Computing net photon flux for binned observations} 
\label{subsec3-1: net phot flux}
In this work, we generate light curves using net photon flux, which is the net count rate per unit area in units of count s$^{-1}$ cm$^{-2}$.
The photon flux is derived from counts measured from the full-band photometry.
We define total counts, $C_{\rm tot}$, as the observed counts within the source aperture and background counts, $C_{\rm bkg}$, as the observed counts within the background region. We scale $C_{\rm bkg}$ by the ratio of areas between the source and background regions to determine the background counts within the source aperture ($C_{\rm bkg, scaled}$), accounting for the different sizes of the source and background regions. Finally, net counts, $C_{\rm net}$, is the difference between the total counts and the scaled background counts. The relation between these parameters is the following: 
\begin{equation}
\label{eqn: counts}
C_{\rm net} = C_{\rm tot} - C_{\rm bkg} \times \frac{A_{\rm src}}{A_{\rm bkg}} = C_{\rm tot} - C_{\rm bkg, scaled}
\end{equation}
\noindent where $A_{\rm src}$ is the area of the source region and $A_{\rm bkg}$ is the area of the background region. We use the inverse of the background scaling parameter (\texttt{backscal}) provided in the {\sc ACIS Extract} data products as the ratio of $A_{\rm src}$ to $A_{\rm bkg}$. 

We divide counts by the product of effective exposure in an epoch, $E_{\rm eff}$, and point spread function (PSF) fraction of the source for the same epoch, $f_{\rm PSF}$, to convert from counts to photon flux. The effective exposure is the product of effective exposure time and effective photon-collecting area of the telescope. 
The PSF is the response function of the telescope to a given point source; thus, the PSF fraction refers to 
the fraction of the counts from a point source enclosed by the detector's aperture.
As an example of the conversion between counts and photon flux,
we convert $C_{\rm net}$ to the observed net photon flux
, $F_{\rm net}$, following the equation below:
\begin{equation}
\label{eqn: net photon flux}
F_{\rm net} = \frac{C_{\rm net}}{f_{\rm PSF}\times{E_{\rm eff}}}
\end{equation}
\noindent 
All three parameters on the right-hand side of Equation \ref{eqn: net photon flux} are obtained from the {\sc ACIS Extract} data products from L17, 
which provide counts, effective exposure, and PSF fraction for each observation. Hence, merging observations into epochs requires either summing or averaging parameters from each observation 
to obtain values 
for an epoch. More specifically, $C_{\rm net}$ is the sum of net counts from all observations that comprise an epoch, while $E_{\rm eff}$ is the sum of the (average) effective exposure from each observation within an epoch,
and $f_{\rm PSF}$ is the weighted average of the PSF fraction from the individual observations making up the epoch. 
As such, each source has five values of $F_{\rm net}$, $C_{\rm net}$, $E_{\rm eff}$, and $f_{\rm PSF}$, one for each epoch.

\subsection{Fitting models to the light curve}
\label{subsec3-2: model}
In order to identify potential TDE candidates, we fit two different models to the light curves of sources in our sample -- a TDE power-law model and a constant model. 
Below we describe the two models and the model fitting procedure.

\subsubsection{The TDE Power-law and the Constant Models}
\label{subsubsec3-2-1: tde and const models}
To model the light curve of a TDE, we adopt a power law with an index of $-5/3$ because a power-law index of $-5/3$ \citep{Rees1988Natur.333..523R} is theoretically expected to describe the decline in X-ray emission for TDEs. We note that in Section \ref{sec5: disc}, we relax this assumption and consider situations in which the net photon flux does not specifically decline as $t^{-5/3}$.
The equation for the TDE power-law model is:
\begin{equation}
\label{eqn: tde mod}
F_{\rm net, model} = (t - t_0)^{-5/3} \times H(t - t_0) \times F_{\rm net, tde}
\end{equation}
\noindent where $t_0$ is the instance when a TDE starts to occur and $H(t - t_0)$ is the Heaviside step function. $H(t - t_0)$ takes the value of unity only when ($t$ $-$ $t_0$) is greater than zero and otherwise $H$($t$ $-$ $t_0$) is zero. 
This model has two free parameters, $t_0$ and $F_{\rm net, tde}$, which is a normalization. 

We set a lower limit for $t_0$ at $T$ = $T_{\rm ep 1} - 10$, 
where $T$ denotes rest-frame time in years and $T_{\rm ep 1}$ is the rest-frame time of the first epoch.
The lower limit on $t_0$ takes into consideration TDEs that may have happened in the past, up to $10$ years before the first epoch. The motivation behind this lower limit is that the decay of the TDE emission often does not last for more than a decade \citep[e.g., see Fig. 2 in][]{Saxton2020SSRv..216...85S}.

We additionally fit the light curves with a constant model. The equation for the constant model is:
\begin{equation}
\label{eqn: cst mod}
F_{\rm net, model} = F_{\rm net, cst}
\end{equation}

\noindent with only one free parameter, $F_{\rm net, cst}$.

\subsubsection{Model Optimization}
\label{subsubsec3-2-3: model opt}
We fit the TDE power-law and constant models to the light curves using \texttt{scipy.optimize} to maximize the log-likelihood.
We calculate the likelihood function for each model following the method detailed here. We use Poisson likelihood, which depends on $C_{\rm tot}$ and model total counts, $C_{\rm tot, model}$. The latter requires us to determine the background photon flux, scaled by the ratio of areas between the source and background regions. The scaled background photon flux is:
\begin{equation}
\label{eqn: bkg phot flux}
F_{\rm bkg, scaled} = \frac{C_{\rm bkg}}{f_{\rm PSF}\times{E_{\rm eff}}} \times \frac{A_{\rm src}}{A_{\rm bkg}} = \frac{C_{\rm bkg, scaled}}{f_{\rm PSF}\times{E_{\rm eff}}}.
\end{equation}

From the models given in Equations \ref{eqn: tde mod} and \ref{eqn: cst mod}, we obtain the model net photon flux, $F_{\rm net, model}$. We then sum $F_{\rm net, model}$ and $F_{\rm bkg, scaled}$ to obtain the model total photon flux, $F_{\rm tot, model}$. Next, we convert $F_{\rm tot, model}$ to $C_{\rm tot, model}$ as follows:
\begin{equation}
\label{eqn: model tot cts}
C_{\rm tot, model} = F_{\rm tot, model} \times (f_{\rm PSF}\times{E_{\rm eff}}).
\end{equation}

We calculate the Poisson probability of the measured counts of each epoch using $C_{\rm tot}$ 
and $C_{\rm tot, model}$ following 
\begin{equation}
\label{eqn: poisson pmf}
P(m, \lambda) = \frac{\lambda^m e^{-\lambda}}{m!}
\end{equation}
\noindent where $m$ is equal to $C_{\rm tot}$ and $\lambda$ is equal to $C_{\rm tot, model}$ for our case. For a given source, the likelihood for each model is the product of the Poisson probability of the five epochs, as shown below: 

\begin{equation}
\label{eqn: poisson likelihood}
L = \prod_{i=1}^{5} P_i
\end{equation}

\noindent where the subscript of $P$ refers to the $i$th epoch.

\begin{figure}[ht!]
\includegraphics[width=\columnwidth]{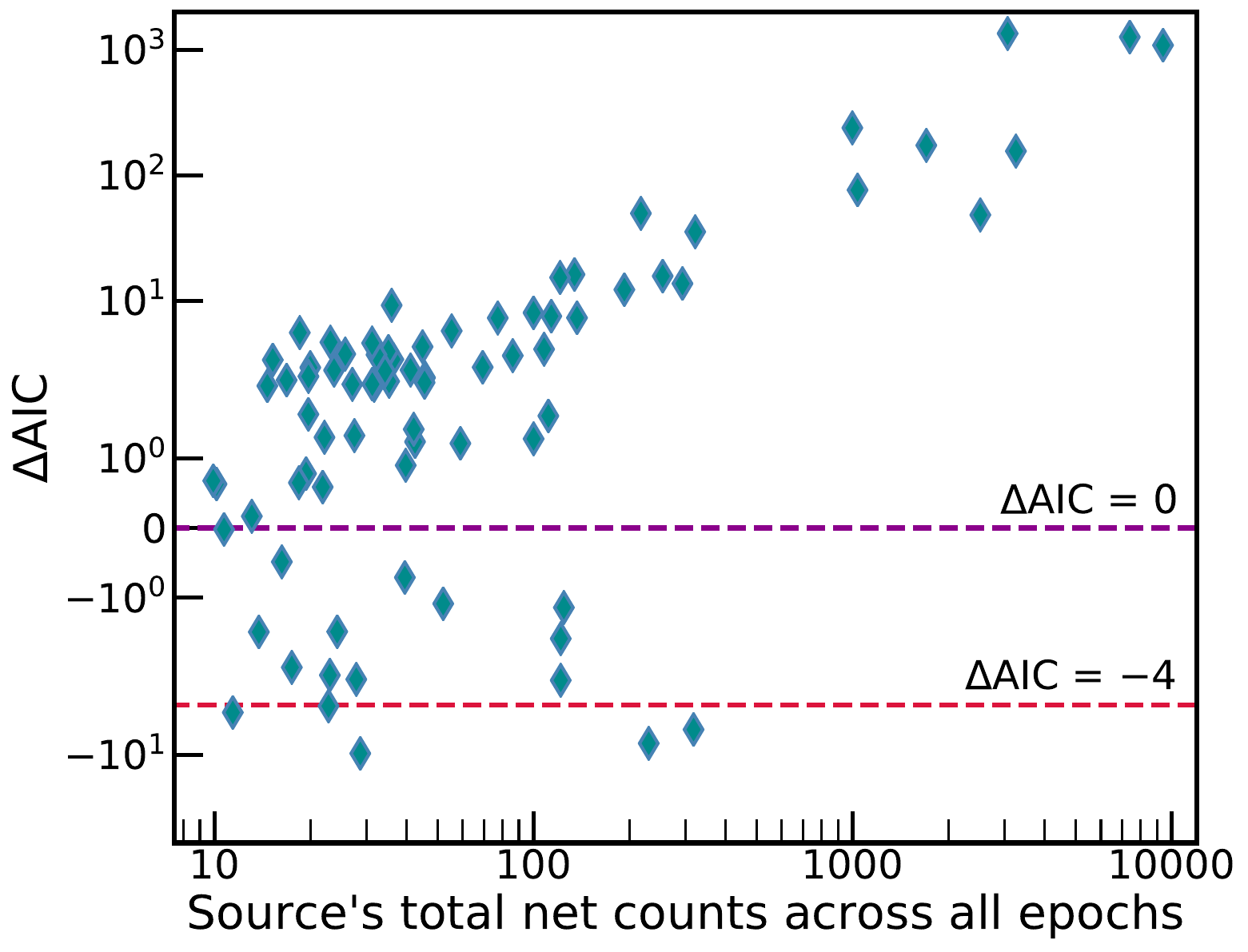}
\caption{\label{fig:aic compare} $\Delta \mathrm{AIC}$ between the best-fit TDE power-law and constant models for the $78$ sources as a function of total observed net counts in the $7$~Ms exposure. The purple dashed line indicates where $\Delta \mathrm{AIC}$ is $0$ and the red dashed line indicates where $\Delta \mathrm{AIC}$ is $-4$. A majority of the sources have positive $\Delta \mathrm{AIC}$, implying that a constant model is preferred over the TDE model.}
\end{figure}

\begin{figure*}[ht!]
\includegraphics[width=\linewidth]{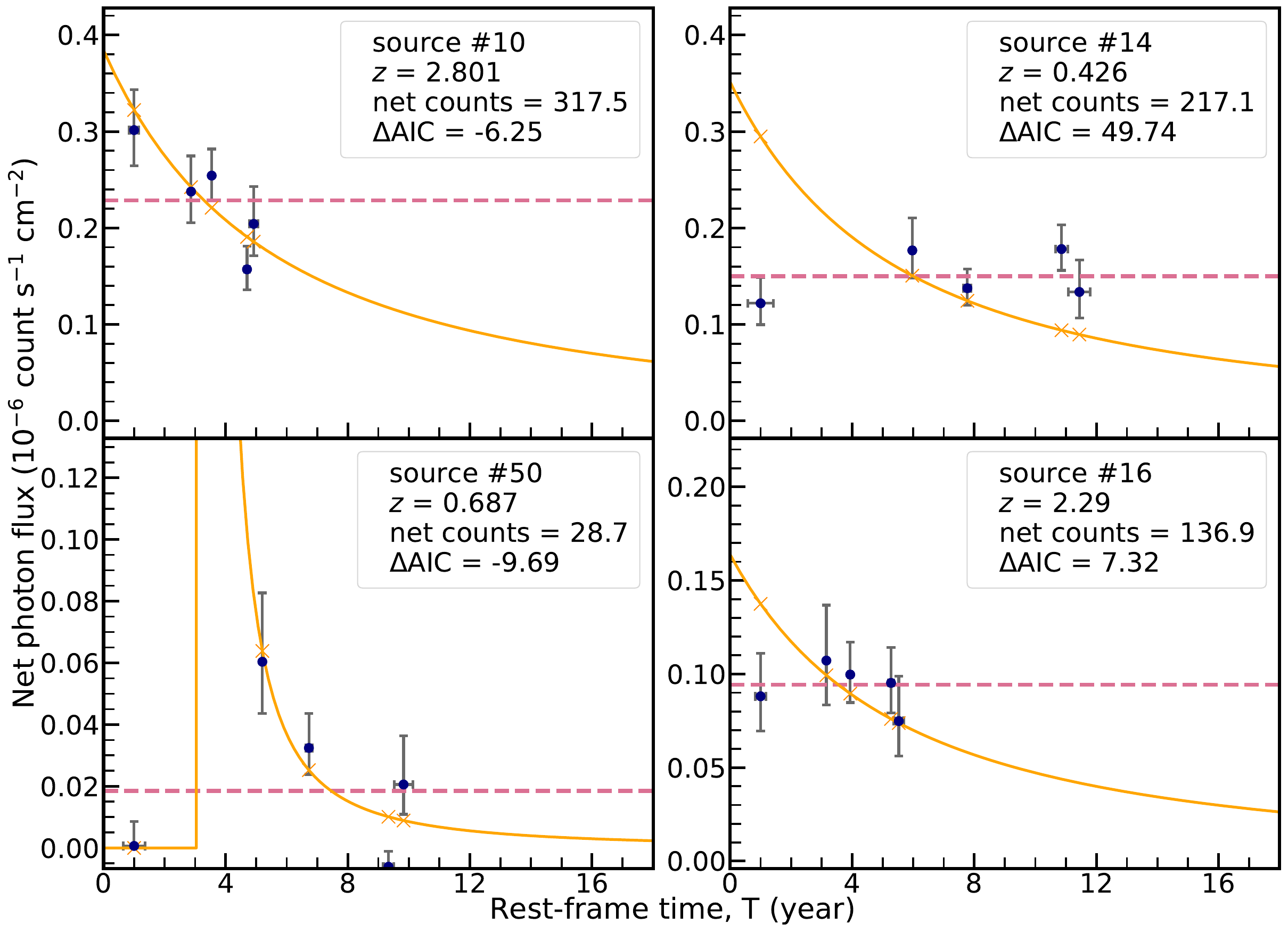}
\caption{\label{fig:light curve} Light curves for four sources in our sample (blue dots), along with the best-fit TDE (orange solid curve) and constant (pink dashed line) models. 
The horizontal error bars indicate the bin width of each epoch, while the rest-frame time of the first epoch is set to year $1$.
The two sources on the left panels are two of the five TDE candidates, while the sources on the right panels are not TDE candidates. The other three TDE candidates not shown here have light curves and best-fit models similar to the candidates presented in the left panels. 
A few source properties are labelled on the panels, along with the $\Delta \mathrm{AIC}$ between the two models. The net counts shown are $C_{\rm net}$ summed over all epochs.}
\end{figure*}

\subsection{Model selection}
\label{subsec3-3: aic}
After fitting both models to the light curves, we apply the Akaike information criterion \citep[AIC;][]{akaikeMR0423716} to compare the quality of the TDE and constant models relative to each other and to select the model with better support from the data.
The AIC value of a model is defined as:
\begin{equation}
\label{eqn: aic}
\mathrm{AIC} = 2k - 2 \ln (L_{\rm max})
\end{equation}

\noindent where $k$ is the number of free parameters and $L_{\rm max}$ is the maximum value of the likelihood function of the fitted model. 
Our focus is on the difference in AIC between our two models, $\Delta \mathrm{AIC}$. 
Following Equation \ref{eqn: aic}, $\Delta \mathrm{AIC}$ is then defined as:
\begin{equation}
\label{eqn: delta aic1}
\Delta \mathrm{AIC} = 2(k_T - k_C) - 2 \ln \bigg( \frac{L_{\rm max, T}}{L_{\rm max, C}} \bigg)
\end{equation}

\noindent where the subscripts $T$ and $C$ refer to the TDE model and the constant model, respectively. The ($k_T - k_C$) term is unity since we have two free parameters for the TDE model and one for the constant model. 
The $\Delta \mathrm{AIC}$ between these two models for each source is 
listed in Table \ref{tab: data}.

A negative value of $\Delta \mathrm{AIC}$ indicates that the TDE model has a smaller value of $\mathrm{AIC}$, and thus, is better supported by the data.
A source with $\Delta \mathrm{AIC}$ of $-4$ and lower indicates that the TDE model is considerably more supported than the constant model \citep{aic}. 
A $\Delta \mathrm{AIC}$ of $-4$ corresponds to the source being $7$ times more likely \citep{aicbic} to be a TDE than not. Accordingly, a lower value of $\Delta \mathrm{AIC}$ implies a higher likelihood of the source being a TDE.
Thus, we impose $\Delta \mathrm{AIC}$ $<$ $-4$ as an empirical criterion to select for TDE candidates and we find a total of $5$ sources that satisfy this empirical threshold.
Figure \ref{fig:aic compare} displays the $\Delta \mathrm{AIC}$ as a function of total observed net counts in the $7$~Ms exposure for all $78$ sources in the sample. As can be seen, the majority of the sources have positive $\Delta \mathrm{AIC}$, indicating that the data from these sources support the constant model better than the TDE model. 
Figure \ref{fig:light curve} shows the light curves of four sources and the best-fit models from the TDE power-law and constant models. These four sources are selected to show sources that prefer the TDE model as well as sources that prefer the constant model. 
In Figure \ref{fig:light curve}, the two left panels show sources with a preference for the TDE model (negative values of $\Delta \mathrm{AIC}$), while the two right panels show sources with a preference for the constant model (positive $\Delta \mathrm{AIC}$ values).

Based on $\Delta \mathrm{AIC}$, there are $5$ TDE candidates in our sample. However, we emphasize that this does not necessarily mean all of them are bona-fide TDEs, as there might be false positives. 
In fact, not all $5$ sources appear to be strong TDE candidates, based on visual inspection of their light curves and best-fit TDE models, as can be seen in the left panels of Figure \ref{fig:light curve}.
As $\Delta \mathrm{AIC}$ is an empirical criterion, it is possible that a non-TDE-related event has been incorrectly selected as a TDE candidate based on its $\Delta \mathrm{AIC}$ value, and vice versa. 
Thus, quantifying the TDE fraction, $f_{\rm TDE}$, as the ratio of the number of TDE candidates to the total number of sources in our sample is not a robust approach.
In Section \ref{sec4: sim}, we estimate the true-positive and false-positive rates for every source in the sample and use them to determine the Bayesian posterior of $f_{\rm TDE}$.

\section{Constraining the TDE Fraction}
\label{sec4: sim}
In order to constrain $f_{\rm TDE}$, we need the true-positive rate and false-positive rate of each source. Here, we define true positive as the case where an actual TDE is categorized as a TDE, and false positive as the case in which the variability of an ordinary AGN is labelled as a TDE. To estimate the true-positive and false-positive rates of each source, we simulate artificial light curves of TDEs (see Section \ref{subsec4-1: sim-tde}) and of ordinary AGNs (see Section \ref{subsec4-2: sim-agn}) that correspond to each source's net counts and redshift.

Using both the simulated TDE and ordinary AGN light curves, we apply a similar series of steps as explained in Section \ref{sec3: analysis} to select for TDE candidates. 
The number of TDE candidates is then used to estimate the true- and false-positive rates for each source.
From these true- and false-positive rates, we use a Bayesian analysis to estimate the 
posterior of the $f_{\rm TDE}$ for our sample (see Section \ref{subsec4-3: posterior}). 
    
\subsection{Simulating TDE light curves}
\label{subsec4-1: sim-tde}
Our goal is to generate a light curve of simulated net photon flux, $F_{\rm net, sim}$. $F_{\rm net, sim}$ can be obtained from the difference between simulated total photon flux, $F_{\rm tot, sim}$, and scaled simulated background photon flux, $F_{\rm bkg, scaled\_sim}$. 
We apply the following series of steps to calculate $F_{\rm tot, sim}$. 
First, for a given low-mass source, we assign to $t_0$ an arbitrary value, ranging from $T_{\rm ep1} - 10$ to $T_{\rm ep 5}$, drawn from a uniform distribution.
With $t_0$, we determine $F_{\rm net, tde}$ using Equation \ref{eqn: tde mod} such that $C_{\rm net, model}$ (converted from $F_{\rm net, model}$ in Equation \ref{eqn: tde mod}) 
summed over the five epochs, equals the $C_{\rm net}$ for a given source, summed over all five epochs.
Thus, we can construct a model light curve ($F_{\rm net, model}$ as a function of $t$) using
the values of the ($t_0$, $F_{\rm net, tde}$) pair. 
Next, following the steps in Section \ref{subsubsec3-2-3: model opt}, we use $F_{\rm net, model}$ and $F_{\rm bkg, scaled}$ to obtain $F_{\rm tot, model}$ and subsequently, $C_{\rm tot, model}$.
We then simulate total counts, $C_{\rm tot, sim}$, from the Poisson distribution, using $C_{\rm tot, model}$ as $\lambda$. We then convert $C_{\rm tot, sim}$ to its flux counterpart, $F_{\rm tot, sim}$, following Equation \ref{eqn: net photon flux}.

\begin{figure*}[ht!]
\includegraphics[width=\linewidth]{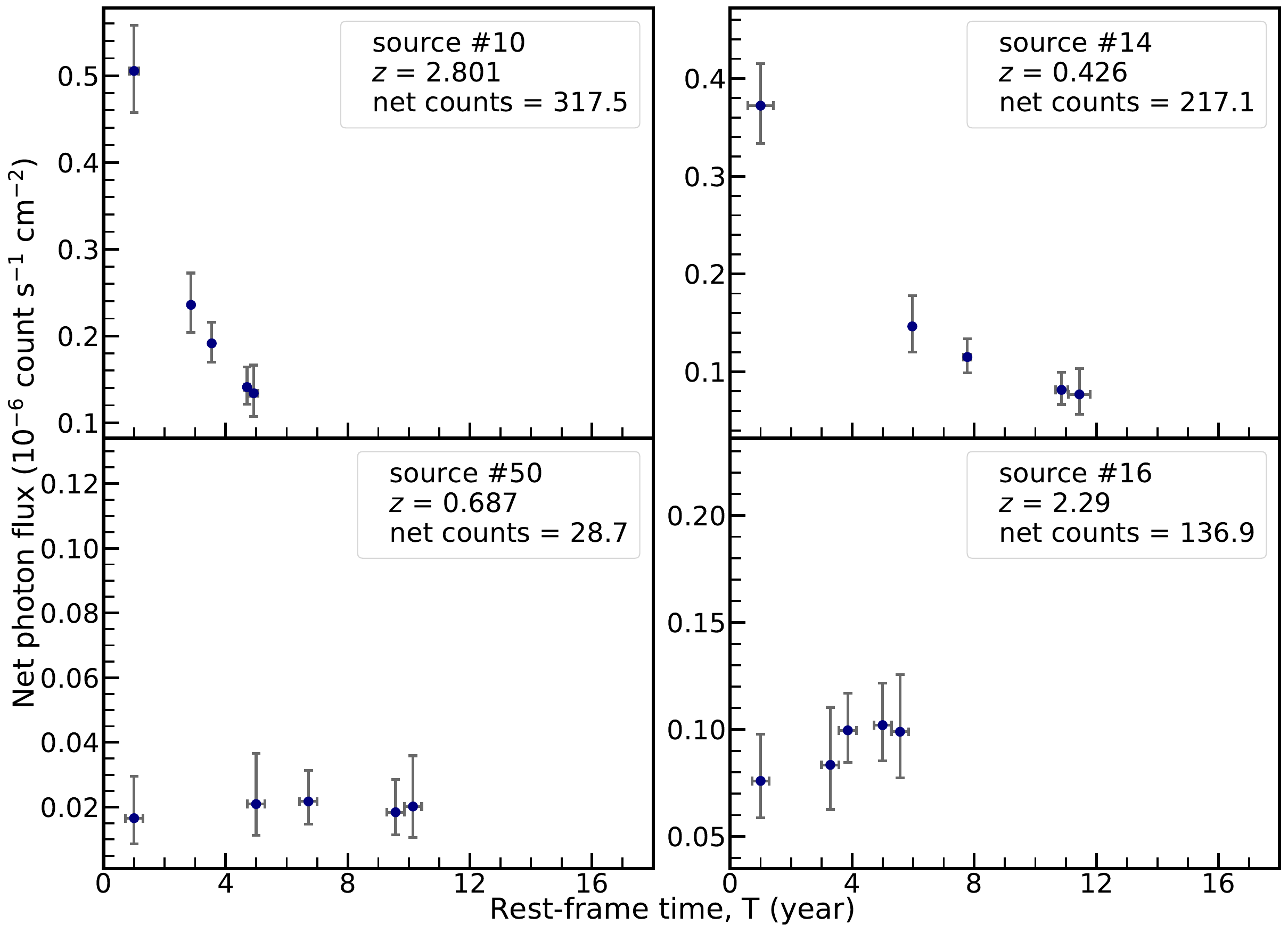}
\caption{\label{fig:sim LCs} The top (bottom) two panels depict the simulated TDE (ordinary AGN) light curves for selected sources in our sample. In all four panels, the blue dots indicate the $5$-epoch data points that form the light curves. The cyan-colored crosses in the bottom panels are the data points in the model ordinary AGN light curve that were not selected to form the simulated ordinary AGN light cruves.
Similar to Figure \ref{fig:light curve}, the horizontal error bars indicate the bin width of each epoch. The rest-frame time of the first epoch is also set to year $1$.
Source properties such as redshift and net counts are labelled on the panels. The net counts shown are $C_{\rm net}$ summed over all epochs. } 
\end{figure*}

Computing $F_{\rm bkg, scaled\_sim}$ is straightforward. We first obtain simulated background counts, $C_{\rm bkg, sim}$, from the Poisson distribution, using $C_{\rm bkg}$ as $\lambda$. Following Equation \ref{eqn: counts}, we then scale $C_{\rm bkg, sim}$ to obtain the scaled simulated background counts, $C_{\rm bkg, scaled\_sim}$. Next, we convert $C_{\rm bkg, scaled\_sim}$ to $F_{\rm bkg, scaled\_sim}$, following Equation \ref{eqn: net photon flux}. Lastly, we subtract $F_{\rm bkg, scaled\_sim}$ from $F_{\rm tot, sim}$ to obtain $F_{\rm net, sim}$, which forms a simulated light curve for the source. The top two panels of Figure \ref{fig:sim LCs} present the simulated TDE light curve for two selected sources.

We now apply the maximum Poisson likelihood estimation as described in Section \ref{subsec3-2: model} to fit both models to the simulated light curve of $F_{\rm net, sim}$. To achieve this, we use $C_{\rm tot, sim}$ as $m$ and $C_{\rm tot, model}$ as $\lambda$ for Equation \ref{eqn: poisson pmf}. Following that, we determine the classification of the source as a TDE candidate or not, using the steps detailed in Section \ref{subsec3-3: aic}. For a given source, we repeat the series of steps as delineated above for a total of $1000$ randomly generated $t_0$. Lastly, we compute the true-positive rate for the source by tallying the number of iterations that identified the source as a TDE candidate and then dividing that tally by $1000$.

The true-positive rate of the $78$ sources in our sample is shown in the top panel of Figure \ref{fig:true false pos}, where it is plotted as a function of total observed net counts in the $7$~Ms exposure. The colored stars refer to the five TDE candidates, while the colored dots are the non-candidates.
The data points are also color-coded by the value of the source's redshift. In general, at similar redshifts, the true-positive rate is higher for sources with higher net counts. 
This is because sources with higher net counts have higher statistical significance, which leads to a higher ability to reach the $\Delta \mathrm{AIC}$ threshold to be considered a TDE candidate.
At similar net counts, the true-positive rate is higher for sources with lower redshifts. Sources at lower redshifts have a longer baseline in the rest frame, which increases the probability of a TDE occurring during the baseline, making it easier to identify TDEs.

\subsection{Simulating ordinary AGN light curves}
\label{subsec4-2: sim-agn}
Next, we generate model light curves of ordinary AGNs using the algorithm described in \citet{Emman2013MNRAS.433..907E}.\footnote{Specifically, we use the {\sc python} implementation by 
\citet{Connolly2016ascl.soft02012C}, available from \href{https://github.com/samconnolly/DELightcurveSimulation}{https://github.com/samconnolly/DELightcurveSimulation}} In order to generate these model ordinary AGN light curves, we need known input parameters for the power spectral density (PSD) and the probability density function (PDF). 
PSD refers to the power of a time series as a function of frequency, while 
PDF refers to the probability distribution of the observed flux. 
For the PSD, we adopt the bending power law from \citet{Paolillo2017MNRAS.471.4398P}:
\begin{equation}
\label{eqn: psd paoli}
\mathrm{PSD}(\nu) = \frac{A}{\nu} \bigg(\frac{1}{1 + \nu/{\nu_b}} \bigg)
\end{equation}
From \citet{Paolillo2017MNRAS.471.4398P}, we adopted a bending frequency, $\nu_b$, of $5.8$ $\times$ $10^{-4}$ s$^{-1}$, and a normalization factor, $A$, of $0.04$ s.

On the other hand, we adopt a lognormal distribution for the PDF. 
We need to estimate the dispersion for the PDF, thus we fit a log-normal distribution to the $F_{\rm net}$ of sources in the L17 catalog with $M_*$ $>$ $10^{10}$ $M_{\odot}$ and sum of $C_{\rm net}$ over five epochs $>$ $500$. The counts cutoff is in place to ensure that the observations have high S/N and statistical significance. 
For the fitting, we focus on X-ray detected massive AGNs that are powered by cold-gas accretion, making them suitable analogs for the ordinary AGNs in the dwarf galaxy sample. We obtain a dispersion of $0.1109$ dex.

Each model light curve has $36$ data points, with an interval of $0.57$ years, making the duration of the entire light curve approximately $20.5$ years. We choose this light-curve duration as it is long enough to cover the rest-frame light curve of all sources. We also use the median bin width of our observed epochs (refer to Table \ref{tab: epoch}) as the interval of our model light curve. 
Next, we want to identify the data points that will be used to form the $5$-epoch ordinary AGN light curves. 
As the model light curve is generated in the rest frame, we converted the observed timestamps for every source from the observed frame to the rest frame. Then, we identify the data point of the model light curve closest to each of the rest-frame timestamp.
Therefore, for each of the model light curve, there are $5$ data points, one for each epoch, similar to the observed light curve.

The model ordinary AGN light curve is in the form of net photon flux, $F_{\rm net, model}$; thus, we use Equation \ref{eqn: net photon flux} to convert $F_{\rm net, model}$ to $C_{\rm net, model}$. We then normalize the light curve by the ratio of the sum of $C_{\rm net, model}$ to the sum of $C_{\rm net}$. This normalization factor adjusts the light curve so that the model net counts correspond to the observed net counts of the source.

We generate $1000$ model ordinary AGN light curves for each source as elaborated above. 
For every light curve, similar to the steps in Section \ref{subsubsec3-2-3: model opt}, we sum $F_{\rm net, model}$ and $F_{\rm bkg, scaled}$ to get $F_{\rm tot, model}$. We then convert $F_{\rm tot, model}$ to $C_{\rm tot, model}$ using Equation \ref{eqn: net photon flux}. Next, we obtain $C_{\rm tot, sim}$ from the Poisson distribution, using $C_{\rm tot, model}$ as $\lambda$. $C_{\rm tot, sim}$ is then converted to $F_{\rm tot, sim}$. 
On the other hand, $F_{\rm bkg, scaled\_sim}$ is calculated following the same steps as delineated in Section \ref{subsec4-1: sim-tde}. Lastly, we obtained $F_{\rm net,  sim}$ by subtracting $F_{\rm bkg, scaled\_sim}$ from $F_{\rm tot, sim}$. We now obtain simulated light curves of ordinary AGNs, $1000$ for each source. In Figure \ref{fig:sim LCs}, the bottom two panels display the simulated ordinary AGN light curves for two sources.

We proceed to fit the TDE and constant models to each of the simulated ordinary AGN light curves using maximum Poisson likelihood estimation as explained in Section \ref{subsec3-2: model},  
and then determine if the source is a TDE candidate, following the steps explained in Section \ref{subsec3-3: aic}.  
Lastly, we compute the false-positive rate from the $1000$ simulated ordinary AGN light curves --- we sum up the number of iterations that classified the source as a TDE candidate, and then divide the sum by $1000$ to obtain the false-positive rate for the source.

\begin{figure}[ht!]
\includegraphics[width=\columnwidth]{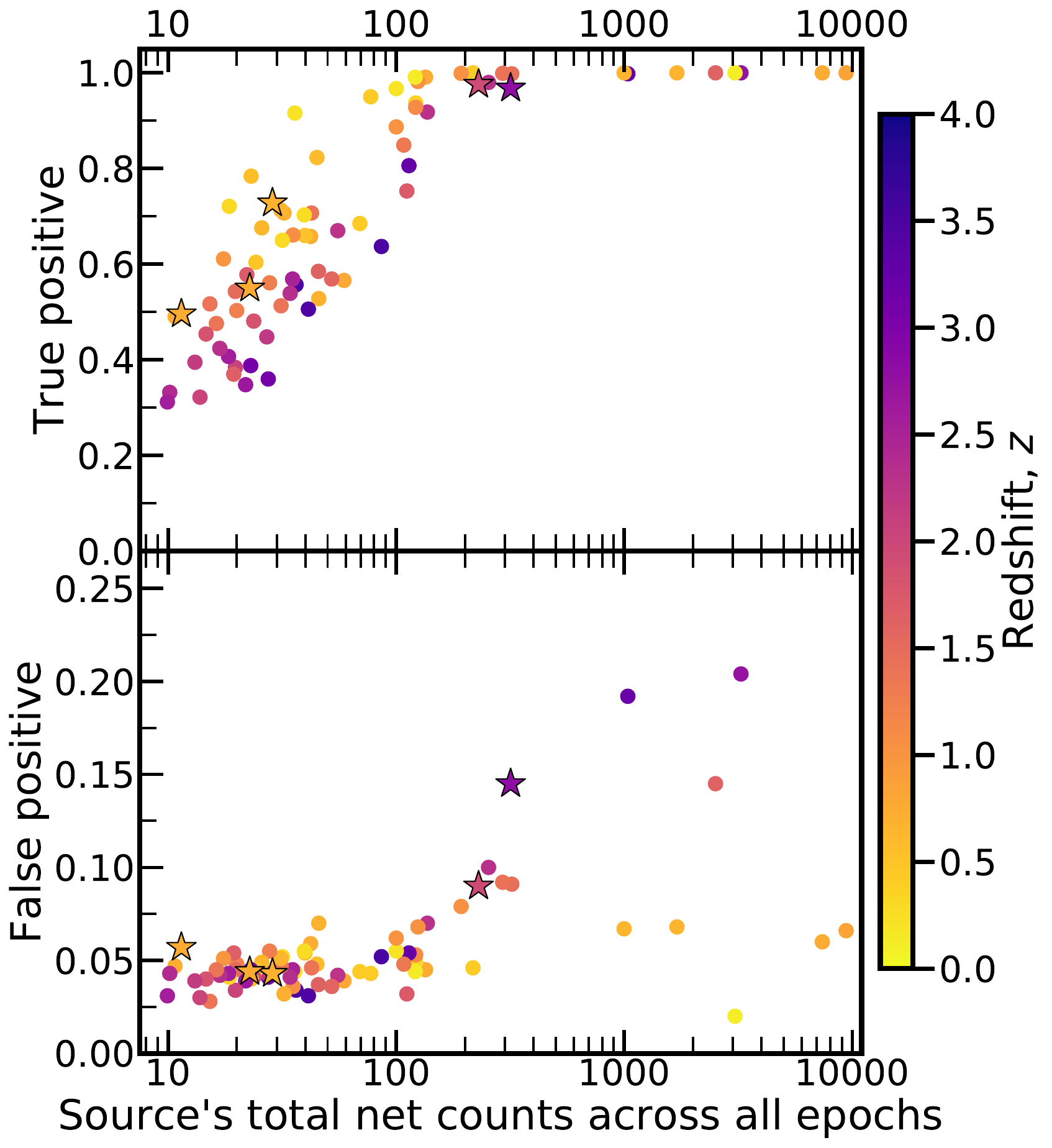}
\caption{\label{fig:true false pos} The top (bottom) panel shows the true (false)-positive rate of the sources as a function of their sum of net counts. Colored stars indicate the TDE candidates in our sample, while the colored dots are non-candidates. 
The data points are color-coded by their redshifts --- yellow tones are used for low redshifts, and purple tones are used for higher redshifts. In general, both true- and false-positive rates increase with net counts at any given redshifts. At a given net count, the true-positive rate is higher for sources with lower redshifts, while the false-positive rate is higher for sources with higher redshifts. Note that the two panels have different limits on the y-axis.}
\end{figure}

The false-positive rate of the sources in our sample is shown in the bottom panel of Figure \ref{fig:true false pos}. 
In general, at similar net counts, the false-positive rate is higher when the source's redshift is higher, although it is only noticeable for sources with high net counts. At lower ($< 100$) net counts, the false-positive rate is similar regardless of redshift.
There is also a noticeably large dispersion in the false-positive rate for high-count sources. 
This is likely due to the higher probability of misidentifying ordinary AGNs as TDEs at higher redshift and shorter observation baseline. 
On the other hand, at any given redshift, the false-positive rate is generally lower for sources with lower net counts. This is similar to the trend of the true-positive rate being lower for sources with lower net counts at similar redshifts. Sources with lower net counts have lower statistical significance, which decreases their ability to meet the $\Delta \mathrm{AIC}$ threshold to be classified as a TDE candidate, thus lowering the false-positive rate.

\subsection{Posterior of the TDE fraction}
\label{subsec4-3: posterior}
Here, using both the true- and false-positive rates, we apply Bayesian analysis to estimate the $f_{\rm TDE}$ posterior from our dataset.
We use Bayes' theorem:
\begin{equation}
\label{eqn: bayes1}
p(\theta|D) = \frac{p(\theta) p(D|\theta)}{p(D)} 
\end{equation}
\noindent where $p(D)$ refers to the evidence, and is a constant independent of data.   
We note that $p(\theta)$ is the prior probability that the model parameter, $\theta$, or in our case $f_{\rm TDE}$, takes on a specific value. For our analysis, we assume a flat prior, therefore, $p(\theta)$, or $p(f_{\rm TDE})$ is assumed to be a constant. This simplifies 
Equation \ref{eqn: bayes1} into:
\begin{equation}
\label{eqn: bayes3}
p(f_{\rm TDE}|D) \propto \frac{p(D|f_{\rm TDE})}{p(D)} 
\end{equation}
\noindent where $p(f_{\rm TDE}|D)$ is the posterior probability distribution, or the probability of getting a certain value of $f_{\rm TDE}$, given the condition that $5/78$ sources in our sample are selected as TDE candidates. 
Lastly, $p(D|f_{\rm TDE})$ refers to the probability of selecting $5/78$ sources as TDE candidates given a $f_{\rm TDE}$.  

Our goal is to compute $p(f_{\rm TDE}|D)$. 
We determine $p(f_{\rm TDE}|D)$ using a Monte Carlo simulation and we first compute $p(D|f_{\rm TDE})$. 
We define the range of $f_{\rm TDE}$ from $0/78$ to $78/78$, at an interval of $1/78$. 
For every $f_{\rm TDE}$, we randomly assign $N_{\rm TDE}$ = $f_{\rm TDE}  \ \times$ $78$ sources as TDEs in the sample. The probability of a source being chosen as a TDE depends on whether it was designated as a TDE or not. If a source is assigned as a TDE, its probability of being selected as a TDE is its true-positive rate. Conversely, if the source is designated as a non-TDE, its probability of being selected as a TDE corresponds to its false-positive rate. 
We apply the following steps to `select' TDEs from our sample. We first draw a random number, ranging from $0$ to $1$, from a uniform distribution. If the number drawn is less than the true-positive rate for a source designated as a TDE (or false-positive rate for an assigned non-TDE), the source is then `selected' randomly as a TDE.
Then, at a given $f_{\rm TDE}$, we run $100000$ Monte Carlo simulations to select TDEs from the sample. We tally the number of iterations for each $f_{\rm TDE}$ that selects exactly $5$ sources as TDEs and divide it by $100000$. This gives us $p(D|f_{\rm TDE})$. 

Then, we divide $p(D|f_{\rm TDE})$ by the total number of iterations with exactly $5$ sources selected as TDEs across all values of $f_{\rm TDE}$. This gives us $p(f_{\rm TDE}|D)$. Finally, we normalize  $p(f_{\rm TDE}|D)$ such that its integral equals to unity, as shown in Equation \ref{eqn: bayes4}:

\begin{equation}
\label{eqn: bayes4}
\int_{-\infty}^{\infty} p(f_{\rm TDE} | D)\ df_{\rm TDE} = 1
\end{equation}

\begin{figure}[ht!]
\includegraphics[width=\columnwidth]{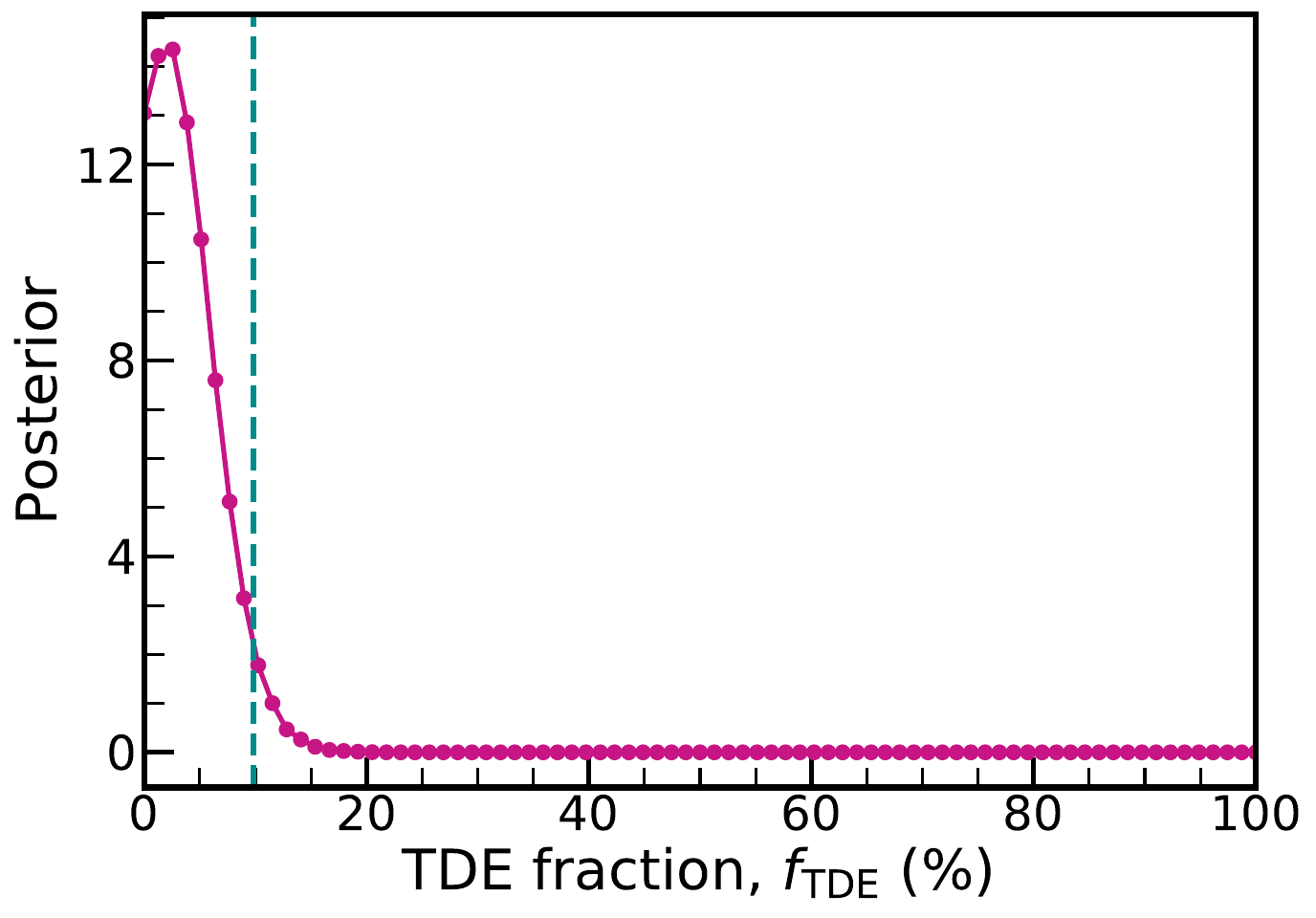}
\caption{\label{fig:posterior}Posterior 
of $f_{\rm TDE}$. The distribution peaks at $f_{\rm TDE}$ = $2.56\%$ and rapidly decreases to $0$. The dashed line marks the $2$-$\sigma$ upper limit of the posterior, which occurs at $f_{\rm TDE}$ = $9.80\%$.}
\end{figure}

Figure \ref{fig:posterior} shows the posterior, i.e. the normalized $p(f_{\rm TDE}|D)$. The peak of the posterior is $2.56\%$, which is very close to zero. We also obtain a $2$-$\sigma$ upper limit of $9.80\%$.

\section{Discussion}
\label{sec5: disc}
In Section \ref{subsec4-3: posterior}, we showed that the posterior $f_{\rm TDE}$ peaks close to zero (2.56\%) with an upper limit of $9.80\%$. 
Thus, most AGNs in dwarf galaxies in our dataset are likely \textit{not} TDE-powered and black-hole growth via the TDE channel may be insignificant. 
However, our previous analyses assume that TDE light curves follow a canonical form of $t^{-5/3}$.
In this section, we supplement our analyses with sanity checks that do not rely on such an assumption.

First, we search for large changes in flux between pairs of epochs for each source. In this way, we are able to identify possible TDEs, which would produce large deviations from a unity flux ratio, without assuming a particular form for the light curve.

\begin{figure*}[ht!]
\includegraphics[width=\linewidth]{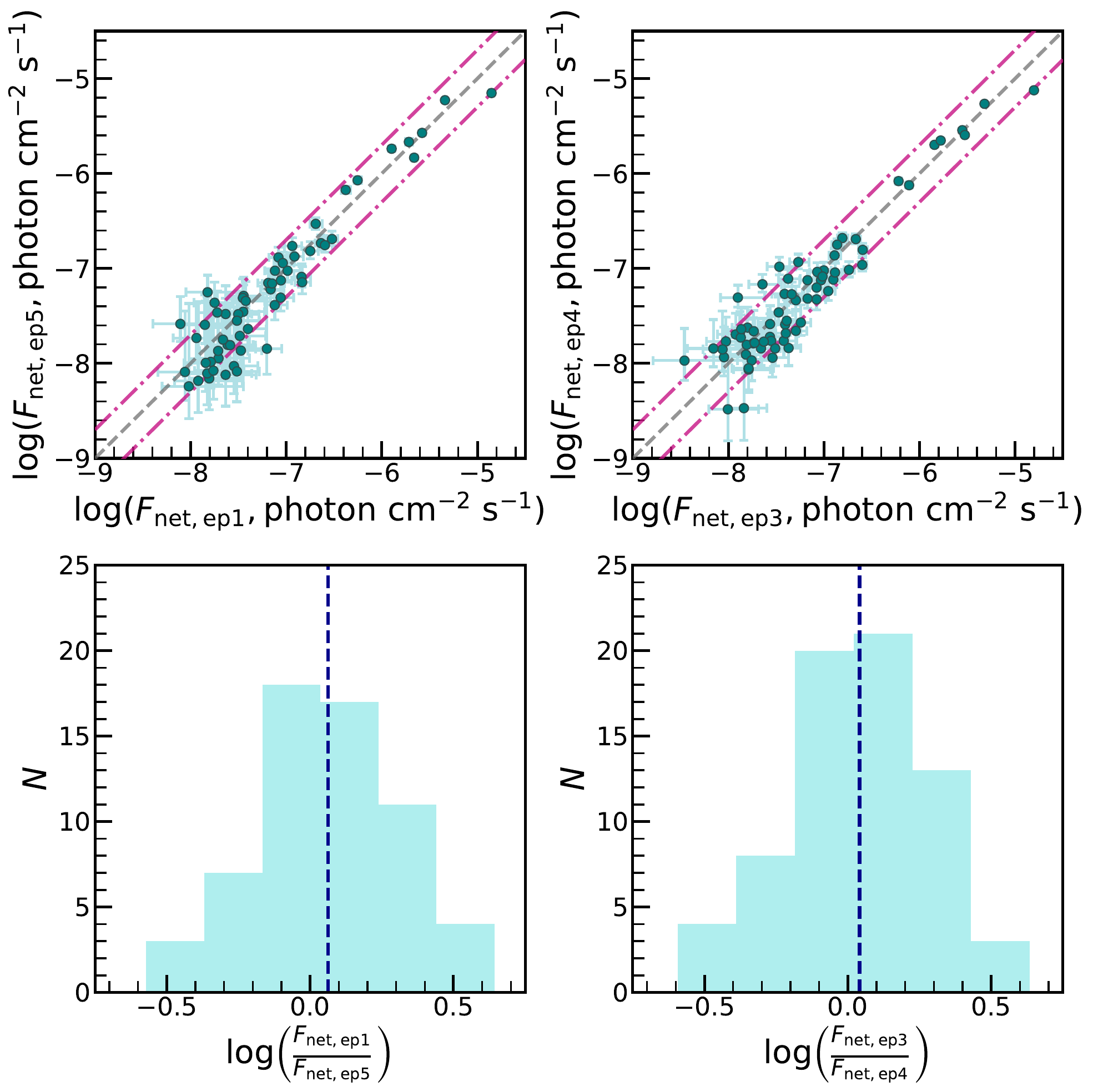}
\caption{\label{fig:log flux ratio} The top two panels show the comparison of full-band net photon fluxes between epochs 1 and 5 (left panel), and between epochs 3 and 4 (right panel). The grey dashed line of unity indicates no change in flux between epochs; the pink dash-dot lines indicate flux changes of a factor of two. 
The fluxes between epochs are similar to within a factor of two for most sources in our sample.
The bottom two panels show the distribution of the log values of the flux ratios between epochs (same pairs as the top panels). The blue dashed vertical line denotes the median of the distribution.
The median log flux ratio and the location of the peak of the distribution are consistent with zero, indicating that there are no systematic changes in flux between epochs.}
\end{figure*}

Figure \ref{fig:log flux ratio} compares $F_{\rm net}$ between epochs 1 and 5 (left column) and between epochs 3 and 4 (right column) as examples. The top two panels each show how the $F_{\rm net}$ of two epochs compare to each other. Nearly all $78$ sources show less than a factor of two change in flux. 
This indicates that there are no extreme changes in flux between epochs (a key signature of TDEs). 
The bottom two panels each show the distribution of the logarithmic values of the $F_{\rm net}$ ratios of two epochs. The median of the distribution, indicated by the blue dashed vertical line, is close to zero. 
The other eight pairs of epochs not shown here show similar results.
The flux vs. flux properties above are consistent with the behavior of ordinary AGNs in general \citep[e.g.,][]{Yang2016ApJ...831..145Y}, disfavoring the claim that our sample is predominantly TDEs.

\begin{figure}[ht!]
\includegraphics[width=\linewidth]
{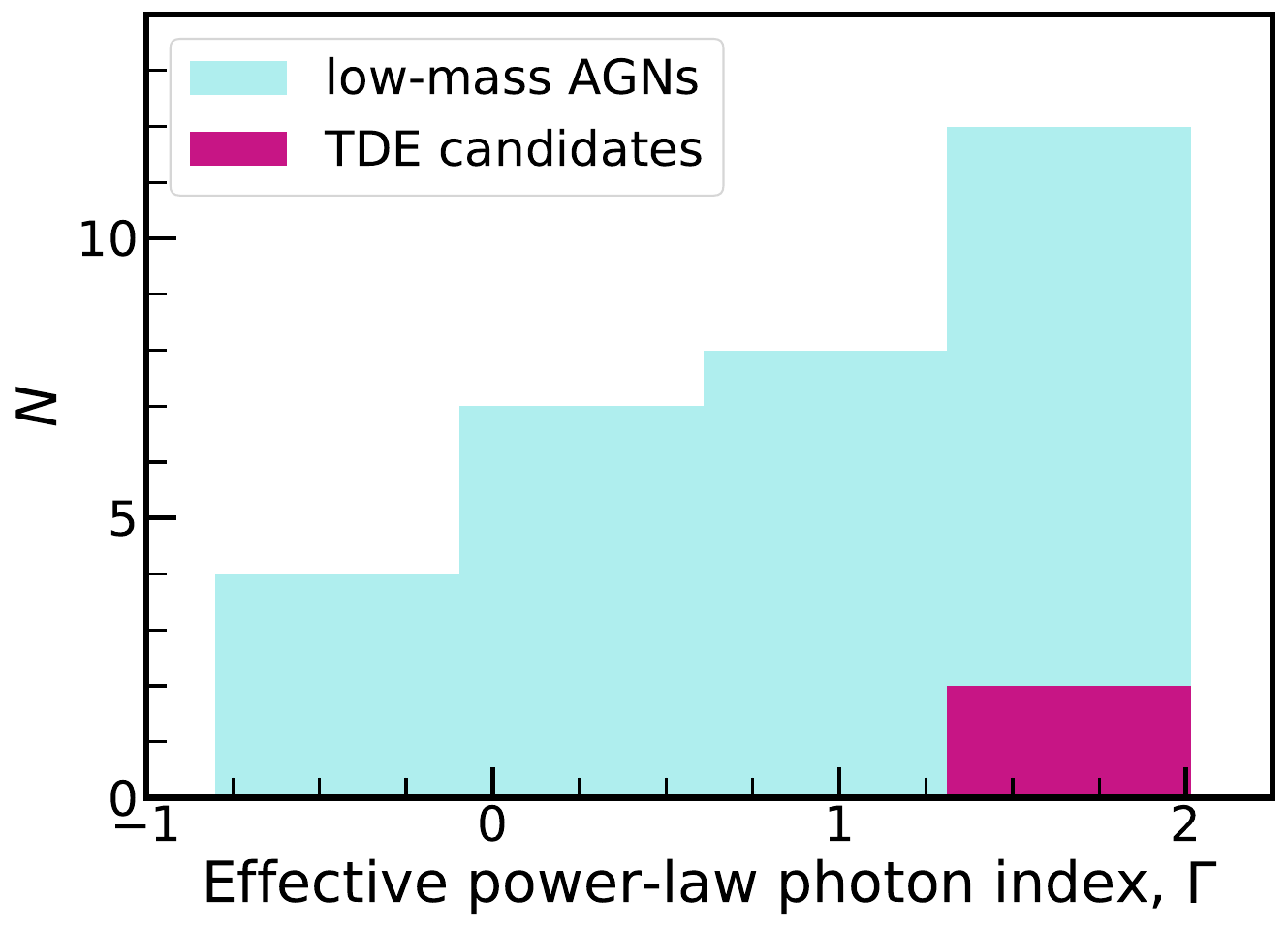}
\caption{\label{fig:gamma dist} Distribution of the effective power-law photon index, $\Gamma$, of sources in our sample. 
Our sources are not TDE-dominated as none of them have $\Gamma$ that is typical of TDEs. The peak of the $\Gamma$ distribution for the TDE candidates (pink histogram) is at $\sim 1.75$, agreeing with the peak of the distribution for the full sample (light blue histogram). The $\Gamma$ distribution for the TDE candidates disagrees with the value of $\Gamma$ typical of TDEs. As such, the TDE candidates are unlikely to be real TDEs.
}
\end{figure}

We use another diagnostic to assess whether our sources are TDE candidates -- we inspect the effective power-law photon index, $\Gamma$, of the sources.\footnote{Admittedly, fitting a more complex model such as a blackbody $+$ power-law model could be more ideal for TDE identification. Unfortunately, our sources generally have limited numbers of counts (median counts $= 40$), preventing us from employing more complex models.} These $\Gamma$ values are adopted from L17. 
This is motivated by the fact that TDEs and ordinary AGNs typically have very different values of $\Gamma$.
Figure \ref{fig:gamma dist} shows the $\Gamma$ distribution of our sample. 
We note that $47$ sources are excluded from the distribution because they are undetected in the soft and/or hard bands, and thus, their values of $\Gamma$ cannot be effectively constrained.
The distribution of $\Gamma$ has a peak between $\Gamma \sim 1.25$ and $\Gamma$=$2$.
This peak disagrees with the typical nature of the soft X-ray emission from a typical (or non-jetted) TDE, which often has $\Gamma \gtrsim $ $4$ \citep[e.g.,][]
{ghosh06, Auchettl2017ApJ...838..149A, Saxton2020SSRv..216...85S, eROSITA_Sazonov2021MNRAS.508.3820S}.\footnote{X-ray emission from TDEs with relativistic jets has a typical $\Gamma$ value of $\sim 1$--$2$.
However, results from previous studies \citep[e.g.,][]{Bower2011ApJ...732L..12B, Bower2013ApJ...763...84B, Brown2015MNRAS.452.4297B} state that only $\sim 10\%$ of TDEs are jetted.}
In contrast, the peak is similar to the X-ray emission from an ordinary AGN, which typically has intrinsic $\Gamma$ $\sim 1.8$ \citep[e.g.,][]{Tozzi2006A&A...451..457T, Marchesi2016ApJ...830..100M, Auchettl2018ApJ...852...37A}.\footnote{We note that 1.8 is the typical intrinsic (without obscuration) value of $\Gamma$ for X-ray emission from ordinary AGNs. However, obscuration can lead to harder X-ray emission, lowering the effective value of $\Gamma$. AGN obscuration and statistical uncertainties are likely the causes for some sources having $\Gamma$ $\leq$ 1.}
Thus, based on their $\Gamma$ values, it is unlikely that the sources in our sample are predominantly TDEs.
Even for the TDE candidates (pink histograms in Figure \ref{fig:gamma dist}) identified in Section \ref{sec3: analysis}, their $\Gamma$ values are all $<3$, suggesting that they are actually false positives.

We emphasize that the two diagnostics performed here are primarily qualitative in nature. 
They serve as a sanity checks, i.e., to strengthen our main conclusion that most of our sources are not TDEs.

Some TDEs could be partial TDEs (PTDEs), and the PTDE decay follows $t^{-9/4}$ instead of $t^{-5/3}$ \citep[e.g.,][]{Coughlin2019ApJ...883L..17C, Miles2020ApJ...899...36M}. Here, we address the concern that fitting a $t^{-5/3}$ model to PTDEs may lead to incorrect conclusions. We simulated PTDE light curves following Section \ref{subsec4-1: sim-tde} by replacing the power-law index of $-5/3$ with $-9/4$. We then fit the $t^{-5/3}$ and constant models to the simulated PTDE light curves to compute the true-positive rates for the sources. We subsequently use the true- and false-positive rates to estimate the $f_{\rm TDE}$ posterior. We find that the posterior is consistent with the posterior calculated in Section \ref{subsec4-3: posterior}. 
The old and new posteriors peak at similar $f_{\rm TDE}$ values (2.56\% and 1.28\%) and have similar $2$-$\sigma$ upper limits (9.80\% and 8.24\%). This finding indicates that our main conclusion remains qualitatively the same (i.e., most of our sources are ordinary AGNs rather than TDEs), even if the TDE decay power-law index is slightly different from $-5/3$.


\section{Summary and future prospects}
\label{sec6: sumz}
In this paper, we systematically search for TDEs among the X-ray detected AGNs in dwarf galaxies from the $7$~Ms CDF-S survey. 
To select dwarf galaxies with $M_* \leq 10^{10} M_{\rm \odot}$, we perform SED fitting for the X-ray sources in the CANDELS region, using {\sc x-cigale} (\S\ref{sec2: data select}). We end up with a sample of $78$ low-mass AGNs.
We extract the $16$-year (observed-frame) light curves for our selected sources (\S\ref{sec3: analysis}). To enhance S/N, the light curves are binned into five epochs, each having $\approx 1$--$2$~Ms exposure. We then fit two models, the TDE power-law model and the constant model, to the light curves. We compare the two models using the AIC statistic, and consider sources with $\Delta \mathrm{AIC} < -4$ as potential TDE candidates. There are a total of $5$ sources with $\Delta \mathrm{AIC}$ $<$ $-4$.

We constrain the TDE fraction by simulating both the TDE and ordinary AGN light curves to obtain true- and false-positive rates for each source (\S\ref{sec4: sim}). We then use these rates and a Bayesian approach to obtain the posterior for the TDE fraction of our observed dataset. From the posterior, we obtain an upper limit of $9.80\%$ for $f_{\rm TDE}$.
This low value of the TDE fraction 
does not support the claim that AGNs in dwarf galaxies are predominantly powered by TDEs \citep[e.g.,][]{Zubovas2019MNRAS.483.1957Z}. 

We compared $F_{\rm net}$ between pairs of epochs to detect large changes in flux and search for possible TDEs in a model-independent way. We found that the changes in flux for most sources are within a factor of two across all pairs of epochs. This small change in flux suggests that the light curves are unlikely the result of a TDE.
We also performed sanity checks based on the flux ratios between epoch pairs and $\Gamma$ values
(\S\ref{sec5: disc}).
These qualitative analyses also suggest that most of our sources are ordinary AGNs rather than TDEs, supporting our main conclusion.

Nevertheless, the sample size of $78$ sources in this work is limited.  
The multi-year surveys from eROSITA and \textit{Athena} will be suitable to probe large samples of sources to better constrain $f_{\rm TDE}$ among AGNs in dwarf galaxies.
\textit{Athena} has a large photon-collecting area, enabling more effective deep surveys. A substantial number of fainter sources could be observed with reasonable S/N, increasing the sample size of sources that would be useful for long-term X-ray variability studies.
eROSITA, thanks to its large field of view, can hunt for TDEs among nearby dwarf galaxies across the entire sky and is expected to. \citet{eROSITA_Khabibullin2014MNRAS.437..327K} predicted that eROSITA could detect up to several thousands of TDE candidates during its all-sky survey, while \citet{eROSITA_Sazonov2021MNRAS.508.3820S} updated the prediction to $\sim$ 700 TDEs. 
On top of that, out of the 13 TDEs from the first \textit{SRG}/eROSITA TDE sample, five of them are hosted by dwarf galaxies (see Appendix C of \citet{eROSITA_Sazonov2021MNRAS.508.3820S} for details on the host galaxies' properties). 
eROSITA's expected yield of AGNs on the order of millions will also be helpful in constraining $f_{\rm TDE}$ in low-mass galaxies.


\acknowledgments
\noindent We thank the referee for helpful feedback that improved this work. 
GY acknowledges funding from the Netherlands Research School for Astronomy (NOVA). WNB acknowledges support from NSF grants AST-2106990 and AST-2407089 and the Penn State Eberly Endowment.
BL acknowledges financial support from the National Natural Science Foundation of China grant 11991053. YQX acknowledges support from the National Natural Science Foundation of China (NSFC-12025303, 11890693). This research has made use of 
\software{Astropy \citep{astropy2013, astropy2018, astropy2022}, Matplotlib \citep{matplotlib}, SciPy \citep{2020SciPy-NMeth}, NumPy \citep{numpy}, {\sc x-cigale} \citep{Boquien2019A&A...622A.103B, Yang2020MNRAS.491..740Y}}
\facility{CXO}

\bibliography{biblio}{}
\bibliographystyle{aasjournal}

\end{CJK*}
\end{document}